\documentclass[11pt]{article}
\usepackage{jheppub2}
\pdfoutput=1
\usepackage{graphicx,epsfig}
\usepackage{float}
\usepackage{subfig}
\usepackage{amsmath,amssymb}
\usepackage[utf8]{inputenc}
\usepackage{hyperref}
\usepackage{caption}
\usepackage{color}
\usepackage{tensor}
\usepackage{mathtools}
\usepackage{physics}
\usepackage{comment}
\usepackage{hyperref}

\newcommand*{\affmark}[1][*]{\textsuperscript{#1}}

\newcommand{\be}{\begin{equation}}

\newcommand{\ee}{\end{equation}}

\title{Semi-classical spacetime thermodynamics}

\author{Ana Alonso-Serrano\affmark[1,]\affmark[2],}
\emailAdd{ana.alonso.serrano@aei.mpg.de}
\author{Marek Li\v{s}ka\affmark[3],}
\emailAdd{liskama4@stp.dias.ie}
\author{and Michał Piotrak\affmark[4]}
\emailAdd{michal.piotrak.23@ucl.ac.uk}

\affiliation{\affmark[1,]Institut für Physik, Humboldt-Universität zu Berlin, Zum Großen Windkanal 6, 12489 Berlin, Germany}
\affiliation{\affmark[2]Max-Planck-Institut f\"ur Gravitationsphysik (Albert-Einstein-Institut), \\Am M\"{u}hlenberg 1, 14476 Potsdam, Germany}
\affiliation{\affmark[3]School of Theoretical Physics, Dublin Institute for Advanced Studies, 10 Burlington Road, Dublin, Ireland}
\affiliation{\affmark[4]Department of Physics and Astronomy, University College London,\\
Gower Street, London, WC1E 6BT, United Kingdom}

\abstract{We derive the semi-classical gravitational dynamics from thermodynamics of local stretched light cones in $2$-dimensional dilaton gravity, explicitly treating the backreaction of quantum matter through the conformal anomaly's effect on the generalized entropy. We also sketch the extension of this analysis to the conformal anomaly in $4$-dimensional semi-classical gravity. In direct connection to this problem, we also tackle the appropriate definition of Wald entropy in thermodynamic derivation of equations of motion for classical scalar-tensor theories. For $2$-dimensional dilaton gravity, we show that the equations of motion follow from Wald entropy associated with dynamical local causal horizons.}

\begin{document}

\maketitle

\section{Introduction}

The entropy-area relation for black hole horizons~\cite{Bekenstein:1972tm,Bekenstein:1973ur,Hawking:1974sw,Hawking:1976de} revealed an interplay between spacetime and thermodynamics. Pushing this insight further, Jacobson showed the full non-linear Einstein's field equations can arise from the Clausius equilibrium relation \cite{Jacobson_1995},
\be Q=T\Delta S\quad \Longrightarrow \quad G_{\mu\nu}+\Lambda g_{\mu\nu}=8\pi G T_{\mu\nu}\;. \label{clausius} \ee
The original argument relies on the observation that locally near any point $p$ in any classical spacetime with a metric $g_{\mu\nu}$, a spacelike foliation can be defined such that through $p$ passes a \emph{local} Rindler horizon with an ascribed uniform Unruh temperature $T$ and entropy \mbox{$S=A/\left(4G\hbar\right)$}\footnote{Unless specified otherwise, we always set $c=k_{\text{B}}=1$, but keep $G$ and $\hbar$ explicit to track the gravitational and quantum effects.}, with $A$ being the area of the spatial cross-section of the horizon. Following the local horizon forward in time, the area is subject to change if matter, characterized by a classical stress-energy tensor $T_{\mu\nu}$, and interpreted as the heat $Q$ that, from the viewpoint of the uniformly accelerating observers perceiving the local Rindler horizon, enters or leaves the horizon. A logical conclusion, then, is that classical gravity emerges from the coarse-grained thermodynamics of some more fundamental theory. Thus began the paradigm of `spacetime thermodynamics'. Since then, this paradigm has been further refined and explored in a number of works \cite{Padmanabhan:2002sha,Padmanabhan:2002jr,Paranjape:2006ca,Padmanabhan:2007en,Parikh:2009qs,Chirco_2010,Baccetti_2013,Jacobson_2012,Parikh_2018,Svesko_2019,Jacobson:2019gco,Alonso:2020,Alonso:2024,Kumar:2023,Kumar:2025}. It has been shown to fully recover the classical equations of motion of any gravitational theory whose Lagrangian can be constructed from just the metric and Riemann tensor (together with a minimally coupled matter Lagrangian). For these more general theories, one needs to replace the entropy proportional to the area with the appropriate Wald entropy \cite{Wald_1993,Iyer_1994}. Furthermore, for technical reasons, Rindler wedges need to be replaced by local compact surfaces, realized either as causal diamonds \cite{Jacobson_2016,Svesko_2019,Alonso:2024} or as stretched light cones \cite{Parikh_2018}. In a parallel development, the Ricci convergence condition has been also recovered from thermodynamics via the second law, even in the presence of 1-loop quantum corrections \cite{Parikh:2015ret,Parikh:2016lys}.

Matter, however, is fundamentally quantum. Accordingly, in an appropriate regime, macroscopic observers live within a semi-classical spacetime governed by the semi-classical Einstein equations, 
\be G_{\mu\nu}+\Lambda g_{\mu\nu}=8\pi G\langle T_{\mu\nu}\rangle\;,\label{eq:semEin}\ee
where classical gravity is sourced by the expectation value of the quantum matter stress-energy tensor in some quantum state. Though semi-classical gravity has a limited range of validity (e.g., for coherent quantum states), it serves as a useful proxy to investigate quantum effects in gravity without having to resort to full fledged quantum gravity. Indeed, horizon thermodynamics historically arose from considering quantum fields on a fixed black hole background \cite{Hawking:1974sw}. The semi-classical field equations (\ref{eq:semEin}) go further by incorporating quantum backreaction effects, i.e., how quantum matter modifies spacetime geometry and vice versa.\footnote{Arguably, outside of Euclidean gravity considerations, backreaction is needed for the thermodynamic interpretation of black hole mechanics. The reason is that, unlike standard black body radiation, Hawking radiation is understood to be excited modes of an external quantum field that is not directly coupled to the background spacetime. With backreaction, where the quantum field sources the background, Hawking radiation knows about the dynamics of the horizon, the temperature of the radiation is identified with the temperature of the horizon, and the black hole evaporates.}

In this article, we derive semi-classical gravitational field equations from thermodynamics of locally constructed stretched light cones, by applying an appropriately modified equilibrium condition
\be \Delta S_{\text{gen}}=0\quad \Longrightarrow \quad \text{semi-classical gravity}\;.\ee
Here, the \emph{classical} gravitational entropy $S$ is replaced by the \emph{generalized} entropy \cite{Bekenstein:1974ax,Gao:2001}, a sum of the gravitational entropy and the von Neuman entropy of the quantum matter fields, schematically $S_{\text{gen}}=S+S_{\text{vN}}^{\text{mat}}+S_{\text{backreaction}}$, where the term $S_{\text{backreaction}}$ accounts for the entropy produced by the backreaction of quantum matter on the classical spacetime. Since all the matter is fundamentally quantum and, hence, accounted for by $S_{\text{gen}}$, there is no classical heat flux term present. This form of the equilibrium condition is reminiscent of the Jacobson's law of the vanishing first order variation of the total entanglement entropy in the spacetime~\cite{Jacobson_2016}
\begin{equation}
\delta S_{\text{EE}}=\delta S_{\text{UV}}+\delta S_{\text{IR}}=0,
\end{equation}
where $\delta S_{\text{UV}}$ and $\delta S_{\text{IR}}$ are the changes of the entanglement entropy due to the variation of th ultraviolet and infrared degrees of freedom, respectively. Then, $\delta S_{\text{UV}}$ diverges with the ultraviolet cutoff squared and scales with the area, and, upon expressing the Newton gravitational constant $G$ in terms of the cutoff, corresponds to the classical gravitational entropy $S$. The second contribution, $\delta S_{\text{IR}}$ is simply the variation of the von Neumann entropy of the matter fields $S_{\text{vN}}^{\text{mat}}$. However, this method has notable limitations. First, it only recovers the linearized semi-classical equations of motion for generic gravitational dynamics~\cite{Bueno_2017} (in the special case of Einstein field equations, the Bianchi identities are powerful enough to recover the full nonlinear dynamics). Second, it does not provide an explicit form of the backreaction terms in the quantum expectation value of the stress-energy tensor. Lastly, $\delta S_{\text{UV}}$ is divergent in the limit of infinite ultraviolet cutoff.

The generalized entropy approach we consider in $2$D represents a significant improvement, providing full non-linear equations with explicit backreaction contributions. Moreover, the generalized entropy has been argued to be manifestly finite and cutoff independent when one accounts for the quantum running of the Newton gravitational constant $G$~\cite{Jacobson:1994,Susskind:1994,Bousso:2015,Marolf:2016,Chandrasekaran_2023,Kudler-Flam:2023,Jensen_2023}.

In the classical limit, the vanishing generalized entropy condition becomes equivalent to the Clausius equilibrium relation employed in the seminal works on thermodynamics of spacetime~\cite{Jacobson_1995,Padmanabhan:2002jr}. To show it, we first note that the quantum backreaction terms in the generalized entropy vanish in the classical limit. The gravitational entropy associated with the horizon remains the same and obeys, in the context of general relativity, $S=A/\left(4G\hbar\right)$. Lastly, two of the authors have shown that the change of the von Neumann entropy of the matter fields $S_{\text{vN}}^{\text{mat}}$ along a spherical horizon in the classical limit approaches the classical heat flux divided by the Unruh temperature~\cite{Alonso:2020} (as the quantum expectation value of the stress-energy tensor in $S_{\text{vN}}^{\text{mat}}$ is replaced by the corresponding classical tensor). Then, the classical limit of the condition $\Delta S_{\text{gen}}=0$ indeed coincides with the Clausius equilibrium relation~\eqref{clausius}.

Our derivation is inspired by Jacobson's original work and by recent developments in the thermodynamics of quantum black holes \cite{Pedraza:2021,Pedraza:2021cvx,Svesko:2022txo,Emparan:2020znc,Panella:2024sor} -- exact black hole solutions to the semi-classical field equations -- where the first law takes the form, $\delta M=T\delta S_{\text{gen}}$ for a black hole mass $M$. 

In fact, the emergence of semi-classical gravity from thermodynamics can be anticipated from the semi-classical first law of horizon thermodynamics. To see it, consider an asymptotically flat black hole surrounded by a fluid. The gravitational first law reads \cite{Bardeen:1973}
\be \delta M=T\delta S_{\text{BH}}+\delta H_{\text{mat}}\;,\ee
where $S_{\text{BH}}=A/4G$ denotes the Bekenstein-Hawking area-entropy, and $\delta H_{\text{mat}}$ encodes the variation in classical stress-energy outside the event horizon. By formally implementing the semi-classical Einstein equations (\ref{eq:semEin}), the classical matter Hamiltonian variation is replaced with its quantum counterpart $\delta\langle H_{\text{mat}}\rangle$. The semi-classical first law then follows upon invoking the first law of entanglement for quantum matter $\delta \langle H_{\text{mat}}\rangle=T\delta S^{\text{mat}}_{\text{vN}}$~\cite{Jacobson:2018ahi,Banihashemi:2023}. By reverse engineering these steps and applying the set-up locally, it is plausible the semi-classical Einstein equations emerge from an appropriate semi-classical first law.

Semi-classical Einstein equations have been derived from thermodynamics in previous works, either considering variations of the total entanglement entropy of the spacetime~\cite{Jacobson_2016,Svesko_2019}, of the Ryu-Takayanagi entropy in the context of the AdS/CFT correspondence~\cite{Faulkner:2014,Faulkner:2017}, and even of the generalized entropy~\cite{Kumar:2023,Kumar:2025}. Moreover, the semi-classical gravitational dynamics has been discussed from a more generic computational perspective~\cite{Pedraza:2022,Pedraza:2022a,Carrasco:2023}. However, an explicit treatment of the backreaction of quantum fields on the spacetime curvature in thermodynamic context is so far missing. Herein, we fill this gap by studying the backreaction problem using the conformal anomaly of the conformal field theories. It is well known that the classical stress-energy tensor of conformally invariant matter is traceless. However, the quantum expectation value of its trace acquires a non-trivial contribution sourced by the spacetime curvature~\cite{Capper:1974,Riegert:1984,Fradkin:1984,Duff:1994} (and, in general, by other fields present, e.g. from the electromagnetic field). Then, an effective gravitational action recovering this anomalous trace of the stress-tensor encodes, at least partially, the backreaction of quantum fields on the spacetime. In general, this method is not sufficient to recover the complete semi-classical backreaction~\cite{Bardeen:2018,Arrechea:2024,Lowe:2025}. However, in $2$D the conformal anomaly action correctly reproduces the complete semi-classical dynamics~\cite{Polyakov:1981,Navarro:1995}. Correspondingly, it has been recently shown that Wald entropy corresponding to the Polyakov semi-classical action for $2$D Jackiw-Teitelboim (JT) gravity~\cite{Teitelboim:1983,Jackiw:1985} accounts for the total generalized entropy, including the backreaction contributions~\cite{Pedraza:2021cvx}. Generalized entropy is in turn equivalent to entanglement entropy~\cite{Kudler-Flam:2023}. Wald entropy then allows us to derive the complete semi-classical equations of JT gravity from thermodynamic equilibrium conditions imposed on local causal horizons. All the aforementioned results apply in the limit of large number of conformal fields coupled to gravity, i.e., large central charge.

A notable advantage of our approach stems from its universality. Entropy proportional to area considered in both seminal Jacobson's papers~\cite{Jacobson_1995,Jacobson_2016} represents a robust prediction for the entanglement entropy of vacuum fluctuations of any quantum field in the presence of a causal horizon~\cite{Sorkin:1986,Srednicki:1993,Solodukhin:2011}. In other words, it makes no reference to gravitational dynamics. Then, using it as an input for deriving the gravitational equations of motion does not rely on any \textit{a priori} knowledge of gravity. This, however, no longer holds for Noetheresque approaches to spacetime thermodynamics, as Wald entropy of a gravitational theory is a Noether charge derived directly from its Lagrangian~\cite{Wald_1993,Iyer_1994}. Then, any thermodynamic derivation of gravitational equations of motion from Wald entropy already implicitly requires the gravitational Lagrangian as an input. Some critics argue that this makes the argument essentially circular. In spite of such criticisms, thermodynamic derivation of the gravitational equations of motion still brings valuable insights. Notably, it shows which theories can be fully reconstructed from their horizon Noether charges. Without further assumptions (see~\cite{Mohd_2016} for a discussion), it is possible only for theories whose Lagrangian can be constructed solely from the metric and the Riemann tensor~\cite{Bueno_2017,Parikh_2018,Svesko_2019}.

The circularity allegations can be disputed by noting that the main claim of spacetime thermodynamics can be presented as a statement of existence of a well-defined entropy functional from which equations of motion can be derived. Then, Wald entropy simply becomes an ansatz which proves existence. Regardless of one's opinion on the circularirty of the Noetheresque approach in general, our particular derivation does not suffer from it thanks to being rooted in the conformal anomaly. Much like area proportionality of entanglement entropy of vacuum fluctuations, conformal anomaly is completely independent of gravitational dynamics. It simply occurs whenever a conformal quantum field are present in a generic curved spacetime~\cite{Duff:1994}. Then,  including  the effect of conformal anomaly in entanglement entropy does not need to invoke gravitational dynamics in any way. This is especially evident in $2$ spacetime dimensions where, as we show, the conformal anomaly contribution to entanglement entropy does not explicitly depend on spacetime curvature nor on the dilaton field (the only dynamical degree of freedom in $2$D gravity). Moreover, entanglement entropy associated with a horizon generically corresponds to Wald entropy~\cite{Solodukhin:2011,Jensen_2023,Kudler-Flam:2023}, allowing us to apply covariant phase space techniques for computing entanglement entropy. In summary, our main results are in this regard equally robust as the derivation of Einstein equations from area-proportional entropy.

The technical challenge we must face lies in applying thermodynamic derivation to a theory with a non-minimally coupled scalar field. The standard definition of Wald entropy is insensitive to the kinetic term of the scalar and cannot serve to recover its contribution to the equations of motion. Moreover, the standard thermodynamic methods fail to recover the equation of motion for the scalar. The issue becomes especially apparent in the context of the $f\left(R\right)$ theories of gravity. On the one hand, their dynamics can be recovered from thermodynamics in a straightforward way. On the other hand, thermodynamics of spacetime apparently fails to recover the equations of motion of the scalar-tensor theory dynamically equivalent to $f\left(R\right)$ gravity. Methods to derive the full equations of motion for scalar-tensor gravity have been proposed in the literature~\cite{Chirco_2011,Mohd_2016}. However, the entropy prescriptions they propose lack a clear physical motivation. In the present work we remedy this issue, showing that for the case of 2D scalar-tensor gravity, the correct entropy is obtained from an improved expression for Wald entropy, recently proposed for dynamical horizons~\cite{Harlow_2020,Margalef:2021,Rignon:2023,Hollands_2024,Visser_2024}. Furthermore, we identify a work term corresponding to the scalar field potential which enters the equilibrium condition and must be included to recover the full equations of motion. We also carefully analyze the peculiar case of $f\left(R\right)$, showing under which assumptions can the dynamics be recovered from the scalar-tensor action.

Upon understanding the thermodynamics of scalar-tensor gravity in the classical setting, we turn to semi-classical dynamics of JT gravity in $2$D. We are able to derive the complete set of equations of motion. We also discuss the possible extension to $4D$ semi-classical gravity. In this case, we face the problem that, as mentioned above, effective actions reproducing the conformal anomaly fail to recover the full semi-classical backreaction. The corresponding Wald entropy then probably does not account for the total generalized entropy. Without the generalized entropy, it becomes unclear how to formulate the correct semi-classical equilibrium condition. Nevertheless, we show that thermodynamics at least succeeds in reproducing the dynamics of the local effective conformal anomaly action, even though the result does not completely capture the semi-classical backreaction.

The paper is organized as follows. In section~\ref{classical}, we first review the construction of the stretched light cones (SLCs) and the derivation of the classical Einstein equations from their thermodynamics. Then, we discuss the recovery of equations of motion for scalar-tensor theories from thermodynamic considerations, generalizing the previously introduced method to work for SLCs and giving it stronger theoretical motivation. Section~\ref{sec:semi2Dgrav} is devoted to the derivation of the full semi-classical equations of JT gravity from generalized entropy. In section~\ref{4D} we sketch the same procedure for $4D$ general relativity, paying attention to the challenges one needs to address in the higher-dimensional setting. Finally, section~\ref{outlook} discusses possible further directions for this semi-classical dynamics program.

\section{Classical spacetime thermodynamics}
\label{classical}

In this section, we first review a method to obtain the classical Einstein equations from thermodynamics of local stretched light cones~\cite{Parikh_2018}. We then extend the derivation to theories of gravity with a non-minimally coupled scalar field, modifying a previously developed approach~\cite{Mohd_2016}. We also provide a physical justification for the choices of the entropy functional made in the original reference. We later employ the tools introduced in this section to derive the semi-classical gravitational equations of motion in $1+1$ spacetime dimensions in section~\ref{sec:semi2Dgrav}.

\subsection{Einstein equations from thermodynamics of stretched light cones}
\label{Einstein from TD}

Herein, we recall the derivation of the Einstein equations from thermodynamics. The stretched light cone geometry \cite{Parikh_2018} we review here will be used throughout the paper.

\paragraph{Stretched light cones.}

Stretched light cones have closed, spherically symmetric spatial cross-sections, which greatly simplifies computing their entropy, both technically and conceptually (see~\cite{Parikh_2018} for details). Let us remark that causal diamonds also have the same advantages, and the results one obtains for stretched light cones and causal diamonds are completely equivalent~\cite{Svesko_2019}, since they simply represent two different ways to explore the thermodynamics of locally constructed light cones.

To describe the geometry of a local stretched light cone in a generic curved spacetime, we can expand the metric around the light cone's apex $P$ in the Riemann normal coordinates~\cite{Brewin_2009}. For a regular spacetime point with coordinates $x^{\mu}$ sufficiently close to $P$, the metric expansion reads
\begin{equation}
g_{\mu\nu}\left(x\right)=\eta_{\mu\nu}-\frac{1}{3}R_{\mu\rho\nu\sigma}\left(P\right)x^{\rho}x^{\sigma}+O\left(x^3\right),
\end{equation}
where $\eta_{\mu\nu}$ denotes the flat spacetime Minkowski metric. The $O\left(x^3\right)$ terms can be neglected since we only consider a small neighborhood of $P$.

A future-oriented light cone with an apex at $P$ is a surface on which the spherical boost generator
\begin{equation}
\xi^{\mu}=\sqrt{x_ix^i}\partial_{t}^{\mu}+\frac{t}{\sqrt{x_ix^i}}x^{j}\partial_{j}^{\mu}=r\partial_{t}^{\mu}+t\partial_{r}^{\mu},
\end{equation}
becomes null, i.e., $t=r$. However, one cannot straightforwardly assign an Unruh temperature to a null wordline, preventing us from computing the corresponding entropy flux. Therefore, we instead introduce a stretched light cone as a timelike surface in the following way. First, we choose a length scale $\alpha$ much smaller than the local curvature length $\lambda_{\text{R}}$ (an inverse of the square root of the largest eigenvalue of the Riemann tensor), $\alpha\ll\lambda_{\text{R}}$. We define the stretched light cone as a timelike hyperboloid $\Sigma$ located just outside the light cone and given by the condition $r^2-t^2=\alpha^2$, where we take $\alpha\ll\lambda_{\text{R}}$. In the limit $\alpha\to0$, $\Sigma$ tends to the light cone. The unit velocity vector tangent to $\Sigma$,
\begin{equation}
\label{velocity}
u^{\mu}=\frac{\xi^{\mu}}{\alpha}\bigg\vert_{r^2-t^2=\alpha^2},
\end{equation}
describes a uniformly accelerating trajectory with an acceleration $a^{\mu}=n^{\mu}/\alpha$, where $n^{\mu}$ denotes the unit, outward-pointing normal to the hyperboloid $\Sigma$
\begin{equation}
\label{normal}
n^{\mu}=\frac{t}{\alpha}\partial_{t}^{\mu}+\frac{r}{\alpha}\partial_{r}^{\mu}.
\end{equation}
The acceleration magnitude equals $1/\alpha$ on $\Sigma$. An observer equipped with a Unruh-de Witt detector following this trajectory then sees the local Minkowski vacuum as a thermal bath of particles at the Unruh temperature
\begin{equation}
\label{Unruh temperature}
T_{\text{U}}=\frac{\hbar a}{2\pi},
\end{equation}
provided that the detector has enough time to thermalize, i.e., that it has been turned on for a time $\tau\gg\alpha$~\cite{Barbado_2012}. In a curved spacetime, one must also take into account that the acceleration changes along the trajectory due to curvature effects. These enter at the order $t/\lambda_{\text{R}}$ and can therefore be neglected for sufficiently small time intervals, $t\ll\lambda_{\text{R}}$.

\paragraph{Entropy of a stretched light cone.}

A radial uniformly accelerating observer whose velocity $u^{\mu}$ obeys equation~\eqref{velocity} cannot access the interior of the light cone. Hence, the light cone represents a causal horizon for such observers in the same sense Rindler horizon does for Cartesian uniformly accelerating observers. Therefore, the radial uniformly accelerating observer perceives entanglement entropy proportional to the area of the spatial cross-section of the light cone at any given time\footnote{An alternative way to understand the association of entropy with a light cone lies in noting that a local light cone can be always understood as a part of a causal diamond. Both entanglement and Wald entropy of causal diamonds have been thoroughly established in the literature, see e.g.~\cite{Jacobson_2019} and references therein.}. The stretched light cone $\Sigma$, being a timelike surface located just outside of the light cone's null boundary has of course larger area than the light cone itself. Since $\Sigma$ is not itself a causal horizon, associating entropy with its area is rather less straightforward. Nevertheless, one may take the area of $\Sigma$ as a good approximation for the area of the light cone, as the former approaches the latter in the limit $\alpha\to0$. Hence, and following the previous works on the subject~\cite{Parikh_2018,Svesko_2019,Alonso:2024,Alonso:2025}, we assume that $\Sigma$ possesses entanglement entropy that is, to the leading order, proportional to the area of its spatial cross-section.

As we show in the following, the relevant thermodynamic characteristics of stretched light cones at time interval $t\in\left(0,\epsilon\right)$, where $\epsilon\ll\alpha$, change only with $\epsilon^2$. Hence, at this time scale, the stretched light cone can be seen as quasi-stationary. Then, we are able to obtain a first law of stretched light cones following, e.g. the covariant phase space algorithm~\cite{Rignon:2023b,Rignon:2023}. On the kinematical level, we can conversely work with the Clausius relation for reversible change of entanglement entropy, which we introduce in the following.

We then have two sources of entropy associated with a stretched light cone $\Sigma$: the entropy of the surface itself and that of the matter flux crossing it. For an equilibrium process, the net change of entropy must be zero, i.e., the changes in entropy from both sources must cancel out together. As we will show, this condition suffices to reconstruct the Einstein equations. The two sources are the following ones:
\begin{itemize}
\item The matter flowing across $\Sigma$ possesses entropy. It can be quantified via the Clausius equilibrium relation, $\Delta S_{\text{C}}=\Delta Q/T_{\text{U}}$, where $T_{\text{U}}$ is the Unruh temperature~\eqref{Unruh temperature}. Evaluating this expression for $\Sigma$ yields~\cite{Jacobson_1995,Baccetti_2013,Parikh_2018}
\begin{equation}
\label{dS Clausius}
\Delta S_{\text{C}}=\frac{2\pi}{\hbar}\int\text{d}^{D-1}\Sigma T_{\mu\nu}\xi^{\nu}n^{\mu}.
\end{equation}
Between times $t=0$ and $t=\epsilon$ Clausius entropy changes by
\be
\Delta S_{\text{C}}=\epsilon^2\frac{\pi\Omega_{D-2}\alpha^{D-2}}{\hbar}T_{tt}.
\ee

Since the evolution of a stretched light cone at short time scales is nearly stationary, we can further compute von Neumann entropy crossing the horizon. For conformally invariant quantum fields in the interior of the stretched light cone at time $t_0\le\epsilon$, von Neumann entropy reads~\cite{Arias:2017}
\be
S_{\text{vN}}\left(t_0\right)=\frac{2\pi}{\hbar}\int_{0}^{\sqrt{\alpha^2+t_0^2}}\text{d}r\int_{\sigma\left(t_0\right)}\text{d}^{D-2}A\langle T_{\mu\nu}\rangle\xi^{\mu}t^{\nu},
\ee
with $\sigma\left(t_0\right)$ being the spatial cross-section of the stretched light cone at $t_0$, and with $\langle T_{\mu\nu}\rangle$ being understood as the quantum expectation value of the stress-energy operator. Since the stress-energy tensor is thermalized, it corresponds to a Kubo-Martin-Schwinger state with respect to the radial boost generator $\xi^{\mu}$. Then, it follows that $T_{\mu\nu}$ is constant under evolution along $\xi^{\mu}$. While the boost generators $\xi^{\mu}$ at times $t=0$ and $t=\epsilon$ differ, their difference is proportional to $\epsilon$ and can be neglected. Then, the change of von Neumann entropy from $t=0$ to $t=\epsilon$ equals
\begin{align}
\nonumber \Delta S_{\text{vN}}=&\frac{2\pi\alpha}{\hbar}\int\langle T_{\mu\nu}\rangle\xi^{\nu}\text{d}\Sigma^{\mu}+O\left(\epsilon^3\right) \\
=&\epsilon^2\frac{\pi\Omega_{D-2}\alpha^{D-2}}{\hbar}\langle T_{tt}\rangle+O\left(\epsilon^3\right). \label{S vN}
\end{align}
Apart from $\langle T_{tt}\rangle$ being a quantum expectation value, this result exactly agrees with $\Delta S_{\text{C}}$. While definitions of Clausius and von Neumann entropy differ, with the former considering entropy flux across the stretched light cone and the latter the entropy contained in its interior, their change over short time periods (where the stretched light cone remains quasi-stationary) is the same. Therefore, the derivation we present in the following can also be phrased in terms of von Neumann entropy of conformal matter fields. We are going to adopt this viewpoint in the semiclassical section~\ref{sec:semi2Dgrav}.

\item The stretched light cone itself possesses entanglement entropy proportional to the area $A$ of orthogonal to $u^{\mu}$, i.e., $S=A/\left(4l_{\text{P}}^2\right)$. At time $t$, this area equals~\cite{Parikh_2018}
\begin{equation}
\nonumber A\left(t\right)\equiv\int_{\sigma\left(t\right)}\text{d}^{D-2}A=\frac{1}{2}\int_{\sigma\left(t\right)}\text{d}\sigma_{\mu\nu}\left(g^{\mu\rho}g^{\nu\sigma}-g^{\mu\sigma}g^{\nu\rho}\right)\nabla_{\rho}\xi_{\sigma},
\end{equation}
where we defined the oriented area element
\begin{equation}
\label{bi-normal}
\text{d}\sigma_{\mu\nu}=\frac{1}{2}\left(n_{\mu}u_{\nu}-u_{\mu}n_{\nu}\right)\text{d}^{D-2}A.
\end{equation}
\end{itemize}

Therefore, the change of entropy of the horizon between times $t=0$ (the light cone's origin) and $t=\epsilon\ll\alpha$ equals\footnote{There exist ambiguities in the definition of a stretched light cone in a generic curved spacetime that affect its area and, hence, entropy~\cite{Alonso:2025}. These ambiguities require further exploration but, for the purpose of this work, we assume that the construction of the stretched light cone is fixed in such a way that equation~\eqref{Delta S} for the area holds.}
\begin{align}
\nonumber \Delta S\equiv&S\left(\epsilon\right)-S\left(0\right) \\
\nonumber =&\frac{1}{4G\hbar}\frac{1}{2}\left(\int_{\sigma\left(\epsilon\right)}\text{d}\sigma_{\mu\nu}\left(g^{\mu\rho}g^{\nu\sigma}-g^{\mu\sigma}g^{\nu\rho}\right)\nabla_{\rho}\xi_{\sigma}-\int_{\sigma\left(0\right)}\text{d}\sigma_{\mu\nu}\left(g^{\mu\rho}g^{\nu\sigma}-g^{\mu\sigma}g^{\nu\rho}\right)\nabla_{\rho}\xi_{\sigma}\right) \\
=&\frac{1}{4G\hbar}\frac{1}{2}\int_{0}^{\epsilon}\text{d}t\int_{\sigma\left(t\right)}\text{d}^{D-2}An_{\mu}\left(g^{\mu\rho}g^{\nu\sigma}-g^{\mu\sigma}g^{\nu\rho}\right)\nabla_{\nu}\nabla_{\rho}\xi_{\sigma}, \label{Delta S}
\end{align}
where we applied the generalized Stokes theorem to obtain the second equality. We can rewrite the second derivative of the spherical boost generator $\xi_{\mu}$ in the following way~\cite{Parikh_2018}
\begin{equation}
\nabla_{\lambda}\nabla_{\nu}\xi_{\mu}=R^{\lambda}_{\;\:\nu\rho\sigma}\xi_{\lambda}+\Phi_{\nu\rho\sigma}.
\end{equation}
If $\xi_{\mu}$ were a Killing vector, $\Phi_{\nu\rho\sigma}$ would vanish by virtue of the Killing identity. However, since the spherical boost generator $\xi^{\mu}$ is not a Killing vector, $\Phi_{\nu\rho\sigma}$ is a complicated expression quantifying the failure of the Killing identity to hold for $\xi_{\mu}$~\cite{Parikh_2018}. After rewriting the second derivative of $\xi_{\mu}$ in this way, the change of entropy $\Delta S$ becomes
\begin{equation}
\Delta S=\frac{1}{4G\hbar}\int_{0}^{\epsilon}\text{d}t\int_{\sigma\left(t\right)}\text{d}^{D-2}A\left(R_{\mu\nu}\xi^{\nu}+\Phi_{\;\:(\mu\nu)}^{\nu}\right)n^{\mu}.
\end{equation}
The integral of the second term yields~\cite{Parikh_2018}
\begin{equation}
\label{irr}
\frac{1}{4l_{\text{P}}^2}\int_{0}^{\epsilon}\text{d}t\int_{\sigma\left(t\right)}\text{d}^{D-2}A\,\Phi_{\;\:(\mu\nu)}^{\nu}n^{\mu}=\frac{\left(D-2\right)\Omega_{D-2}}{8l_{\text{P}}^2}\alpha^{D-4}\epsilon^2+O\left(\alpha^{D-3}\right).
\end{equation}
This term is non-vanishing even in flat spacetime with no Clausius entropy flux present to compensate for it. Therefore, it cannot correspond to an equilibrium process. Rather, it is driven by the flat spacetime expansion of the wordlines generating $\Sigma$. This expansion can be thought of as an irreversible process~\cite{Parikh_2018}, akin to a free expansion of an ideal gas, which also increases its entropy without any heat flux being present. This irreversible entropy production then needs to be subtracted before applying the equilibrium Clausius relation corresponding to a reversible process. The reversible change in entropy thus equals
\begin{equation}
\label{dS rev}
\Delta S_{\text{rev}}=\frac{1}{4G\hbar}\int_{0}^{\epsilon}\text{d}t\int_{\sigma\left(t\right)}\text{d}^{D-2}AR_{\mu\nu}\xi^{\nu}n^{\mu}.
\end{equation}

\paragraph{Einstein equations.}

We are now ready to apply the Clausius equilibrium relation
\begin{equation}
\Delta S_{\text{rev}}=\Delta Q/T_{\text{U}}=\Delta S_{\text{C}}.
\end{equation}
Plugging in the expressions for the reversible change of Wald entropy~\eqref{dS rev} and for the Clausius matter entropy flux~\eqref{dS Clausius}, we obtain
\begin{equation}
\label{equilibrium condition}
\int_{0}^{\epsilon}\text{d}t\int_{\sigma\left(t\right)}\text{d}^{D-2}A\left(\frac{1}{4G\hbar}R_{\mu\nu}-\frac{2\pi}{\hbar}T_{\mu\nu}\right)\xi^{\nu}n^{\mu}=0.
\end{equation}
The boost generator $\xi^{\nu}$ and the normal vector $n^{\mu}$ are arbitrary in the sense that we can construct a stretched light cone with an apex at $P$ for any choice of the timelike vector $\xi^{\nu}$ tangent to it. Then, equation~\eqref{equilibrium condition} implies that the expression contracted with $\xi^{\nu}n^{\mu}$ must vanish at $P$, up to a term proportional to the metric tensor, since $g_{\mu\nu}\xi^{\nu}n^{\mu}=0$ (the vectors are by construction orthogonal). Therefore, it holds
\begin{equation}
\label{Einstein precursor}
R_{\mu\nu}-8\pi G T_{\mu\nu}+\Psi g_{\mu\nu}=0.
\end{equation}
To determine $\Psi$, we can take the divergence of the equation, apply the contracted Bianchi identities and invoke the local stress-energy conservation condition, $\nabla^{\nu}T_{\mu\nu}=0$. We obtain
\begin{equation}
\Psi=-\frac{1}{2}R+\Lambda,
\end{equation}
where $\Lambda$ is an arbitrary integration constant. Equation~\eqref{Einstein precursor} then becomes
\begin{equation}
R_{\mu\nu}-\frac{1}{2}Rg_{\mu\nu}+\Lambda g_{\mu\nu}=8\pi G T_{\mu\nu}.
\end{equation}
We have recovered the Einstein equations, with the integration constant $\Lambda$ playing the role of the cosmological constant. Although the equations were derived at an arbitrary point $P$, the strong equivalence principle guarantees their validity throughout the spacetime~\cite{Chirco_2010}.

\subsection{Non-minimally coupled gravity}
\label{subsec:n-m-g}

The thermodynamic derivation we reviewed in the previous section works not just for general relativity, but also for a class of modified theories of gravity whose Lagrangian can be written as a function of the metric and the Riemann tensor~\cite{Jacobson_2012,Bueno_2017,Parikh_2018,Svesko_2019,Alonso:2024}. In this case, the entropy of the horizon is no longer directly proportional to its area. Instead, it must be generalized to the appropriate Wald entropy, which for stationary black holes corresponds to the conserved Noether charge of the Killing vector tangent to the horizon~\cite{Wald_1993} (see appendix~\ref{Wald entropy review} for details). For a stretched light cone, it has been argued to be analogously proportional to the bulk Noether charge associated with the boost generator $\xi^{\mu}$. Wald entropy then reads
\begin{equation}
\label{SW}
S_{\text{W}}=\frac{2\pi}{\hbar\kappa}\int_{\mathcal{H}}Q^{\nu\mu}_{\xi}\text{d}\sigma_{\mu\nu},
\end{equation}
where $Q^{\nu\mu}_{\xi}$ denotes the bulk Noether charge antisymmetric tensor density, the integration is carried out over the cross-section of the stretched light cone orthogonal to $\xi^{\mu}$ and the oriented area element $\text{d}\sigma_{\mu\nu}$ is given by equation~\eqref{bi-normal}. For Lagrangians constructed from the metric and Riemann tensor, $Q^{\nu\mu}_{\xi}$ equals
\begin{equation}
\label{charge f(g,R)}
Q^{\nu\mu}_{\xi}=-2P^{\mu\nu\rho\sigma}\nabla_{\rho}\xi_{\sigma}+4\nabla_{\rho}P^{\mu\nu\rho\sigma}\xi_{\sigma},
\end{equation}
where
\begin{equation}
\label{P tensor}
P^{\mu\nu\rho\sigma}=\frac{\partial \mathcal{L}}{\partial R_{\mu\nu\rho\sigma}}.
\end{equation}
Thermodynamics of stretched light cones allows us to straightforwardly derive the corresponding equations governing the gravitational dynamics~\cite{Parikh_2018} from any Noether charge of the form~\eqref{charge f(g,R)}.

However, it fails in two important cases~\cite{Mohd_2016}. First, if the gravitational Lagrangian depends on derivatives of the Riemann tensor, they yield contributions to equations of motion that cannot be recovered from Wald entropy. Second, for theories with non-minimal coupling (e.g. scalar-tensor theories of gravity), Wald entropy does not capture the contributions to the equations of motion coming from the terms independent of the Riemann tensor. Moreover, it fails to recover the equations of motion for the additional non-minimally coupled fields, leading to an incomplete description of the gravitational dynamics. Since a thermodynamic study of one such a non-minimally coupled model is the main goal of this paper, we need a way to remedy this shortcoming. Rather than addressing the situation in general, we sketch here the problem and its resolution on the example of the scalar-tensor reformulation of $f\left(R\right)$ theories of gravity.

\paragraph{Thermodynamics of stretched light cones and $f\left(R\right)$ gravity.}

The action for any $f\left(R\right)$ theory reads
\begin{equation}
I=\int f\left(R\right)\sqrt{-\mathfrak{g}}\text{d}^{n}x+I_{\text{m}},
\end{equation}
where $f\left(R\right)$ is some function of the scalar curvature and $I_{\text{m}}$ denotes the action for (minimally coupled) matter fields. The corresponding Wald entropy associated with a stretched light-cone is given by equation~\eqref{SW}, where we have for the Noether charge
\begin{equation}
Q^{\nu\mu}_{\xi}=\frac{1}{16\pi G}f'\left(R\right)\left(g^{\mu\rho}g^{\nu\sigma}-g^{\mu\sigma}g^{\nu\rho}\right)\nabla_{\rho}\xi_{\sigma}-\frac{1}{8\pi G}\left(g^{\mu\rho}g^{\nu\sigma}-g^{\mu\sigma}g^{\nu\rho}\right)\xi_{\sigma}\nabla_{\rho}f'\left(R\right).
\end{equation}
For the change of Wald entropy over a slice of the stretched light cone we obtain~\cite{Parikh_2018}, in an analogous way as in the case of general relativity,
\begin{align}
\nonumber \Delta S_{\text{W}}=&\frac{1}{4G\hbar}\int_{0}^{\epsilon}\text{d}t\int_{\sigma\left(t\right)}\text{d}^{D-2}\mathcal{A}\Big[\left(f'R_{\mu\nu}-\nabla_{\mu}\nabla_{\nu}f'+g_{\mu\nu}\nabla^{\lambda}\nabla_{\lambda}f'\right)\xi^{\nu} \\
&+f'\Phi_{\;\:(\mu\nu)}^{\nu}+\nabla^{\nu}f'\nabla_{(\nu}\xi_{\mu)}-\nabla_{\mu}f'\nabla_{\nu}\xi^{\nu}\Big]n^{\mu}.
\end{align}
It has been shown that the second line partially corresponds to the irreversible entropy production and the remaining terms can be removed by adding suitable subleading terms to the definition of the boost generator $\xi^{\mu}$~\cite{Parikh_2018} (higher order corrections to this vector are in principle unfixed). Only the first line then yields the reversible change of Wald entropy and equating it to the Clausius entropy flux~\eqref{dS Clausius} implies
\begin{equation}
f'R_{\mu\nu}-\nabla_{\mu}\nabla_{\nu}f'+\Psi g_{\mu\nu}=8\pi G T_{\mu\nu}.
\end{equation}
Taking a divergence of this equation and using that $\nabla_{\nu}T_{\mu}^{\;\:\nu}=0$ finally fixes the arbitrary function $\Psi$ to be
\begin{equation}
\Psi=-\frac{1}{2}f+\Lambda,
\end{equation}
with $\Lambda$ being an arbitrary integration constant. In this way, we finally obtain the equations of motion for $f\left(R\right)$ gravity
\begin{equation}
f'\left(R\right)R_{\mu\nu}-\frac{1}{2}fg_{\mu\nu}-\nabla_{\mu}\nabla_{\nu}f'\left(R\right)+g_{\mu\nu}\nabla^{\lambda}\nabla_{\lambda}f'\left(R\right)=8\pi GT_{\mu\nu}.
\end{equation}
Thence, we are able to fully recover the equations of motion for any $f\left(R\right)$ theory from its Wald entropy. The same is true for the entire class of gravitational theories whose Lagrangian can be constructed just from the metric and the Riemann tensor~\cite{Jacobson_2012,Parikh:2009qs,Bueno_2017,Parikh_2018,Svesko_2019}.

\paragraph{Scalar-tensor form of $f\left(R\right)$ gravity.}

Provided that $f''\ne 0$, we can rewrite $f\left(R\right)$ as a scalar-tensor theory with the action~\cite{Bueno_2016}
\begin{equation}
\label{f(R) scalar}
I=\frac{1}{16\pi G}\int\left(\phi R-V\left(\phi\right)+L_{\text{m}}\right)\sqrt{-\mathfrak{g}}\text{d}^{D}x,
\end{equation}
where
\begin{equation}
V\left(\phi\right)=\phi R\left(\phi\right)-f\left(R\left(\phi\right)\right).
\end{equation}
The equations of motion read
\begin{align}
R'\left(\phi\right)\left[\phi-f'\left(R\left(\phi\right)\right)\right]&=0, \label{fr eom 1} \\
\phi R_{\mu\nu}-\nabla_{\mu}\nabla_{\nu}\phi+\left[\nabla_{\lambda}\nabla^{\lambda}\phi-\frac{1}{2}f\left(R\left(\phi\right)\right)\right]g_{\mu\nu}&=8\pi GT_{\mu\nu}. \label{fR eom 2}
\end{align}
The first equation fixes $\phi=f'\left(R\right)$ and the second one is then equivalent to the equations of motion of $f\left(R\right)$ gravity.

Oddly, this scalar-tensor formulation of $f\left(R\right)$ gravity does not allow a straightforward derivation of the equations of motion from thermodynamics, even though we have just shown it to be possible with the standard $f\left(R\right)$ action. The reason is that while the reversible Clausius relation implies traceless part of equation~\eqref{fR eom 2}, it fails to provide equation~\eqref{fr eom 1}, which connects the scalar field $\phi$ with the curvature. It also does not offer any way to recover the information about the scalar field potential from the traceless form of equation~\eqref{fR eom 2}. However, an alternative thermodynamic approach able to address (among other cases) the non-minimal coupling has been proposed~\cite{Mohd_2016} and we will show that it allows us to recover both equations\footnote{An earlier approach applicable to Brans-Dicke theories which relies on fixing the initial expansion of the horizon should also be noted~\cite{Chirco_2011}.}. It relies on ambiguities in the definition of Wald entropy, which can be amended by certain terms without affecting the well established results for stationary black holes. As we discuss in more detail in appendix~\ref{Wald entropy review}, such changes are actually necessary for the case of non-stationary black hole spacetimes. Otherwise, the generalized second law cannot be satisfied~\cite{Dong_2014,Wall_2015,Hollands_2024,Visser_2024}.

The augmented Wald entropy expression able to recover the equations of motion reads for null Killing horizons~\cite{Mohd_2016}
\begin{equation}
\label{S aug}
S=S_{\text{W}}+2\int M^{\lambda[\mu}\xi^{\nu]}\xi_{\lambda}\text{d}\sigma_{\mu\nu},
\end{equation}
where the integration is carried out over a spatial slice of the horizon and $M^{\lambda\mu}$ is a symmetric tensor. It is easy to check that such a contribution indeed vanishes for stationary black holes\footnote{Assuming that $M^{\mu\lambda}$ does not diverge on the horizon's bifurcation surface~\cite{Jacobson_1994,Sarkar_2011,Gadioux_2023}.}. The ansatz proposed for the tensor $M^{\lambda\mu}$ reads, for theories whose Lagrangian $L$ does not contain derivatives of the Riemann tensor~\cite{Mohd_2016},
\begin{equation}
\label{M tensor}
M^{\lambda\mu}=\frac{\partial \mathcal{L}}{\partial g_{\lambda\mu}}+2P^{\rho\sigma\tau(\mu}R_{\rho\sigma\tau}^{\quad\;\:\lambda)}.
\end{equation}

The original augmented entropy~\eqref{S aug} has been proposed for null, (approximately) Killing horizons. Then, relying on their properties, the contribution to the change of entropy between two horizon slices reduces to~\cite{Mohd_2016}
\begin{equation}
\label{DS aug}
\Delta S=\Delta S_{\text{W}}+2\int\nabla_{\mu}\left(M^{\lambda[\mu}\xi^{\nu]}\xi_{\lambda}\right)k_{\nu}\text{d}^3\Sigma=\Delta S_{\text{W}}-\int M_{\mu\nu}\xi^{\mu}k^{\nu}\text{d}^3\Sigma,
\end{equation}
where $k^{\nu}$ is a null normal to the horizon collinear with the (approximate) Killing vector $\xi^{\mu}$. For a timelike surface, such as a stretched light cone, the normal and the vector $\xi^{\mu}$ tangent to the surface are no longer collinear with one another and equation~\eqref{DS aug} no longer holds for entropy~\eqref{S aug}. Nevertheless, we can find a modified version of the augmented entropy prescription~\eqref{S aug} that reproduces the desired entropy change~\eqref{DS aug} for stretched light cones. The prescription reads
\begin{equation}
\label{entropy SLC}
S_{\text{SLC}}=S_{\text{W}}+\frac{2}{D+1}\int M^{\lambda[\mu}\alpha n^{\nu]}\xi_{\lambda}\text{d}\sigma_{\mu\nu}.
\end{equation}
The divergence of the additional term contracted with $n_{\nu}$ equals
\begin{align}
\nonumber \frac{2}{D+1}\nabla_{\mu}\left(M^{\lambda[\mu}\alpha n^{\nu]}\xi_{\lambda}\right)n_{\nu}=&\frac{1}{D+1}\Bigg[n_{\nu}\nabla_{\mu}\left(\alpha M^{\lambda[\mu}\right)n^{\nu]}\xi_{\lambda}+\alpha n_{\nu} M^{\lambda\mu}\xi_{\lambda}\nabla_{\mu}n^{\nu} \\
&-\alpha n_{\nu} M^{\lambda\nu}\xi_{\lambda}\nabla_{\mu}n^{\mu}+\alpha n_{\nu} M^{\lambda\mu}n^{\nu}\nabla_{\mu}\xi_{\lambda}-\alpha n_{\nu} M^{\lambda\nu}n^{\mu}\nabla_{\mu}\xi_{\lambda}\Bigg].
\end{align}
The first term is $O\left(\alpha^2\right)$ and can be neglected, the second one vanishes since $n^{\nu}$ is normalized to unity. The fourth term is $O\left(t\right)$ and can be discarded since $t\le\epsilon\ll\alpha$. The remaining two terms together yield $-\left(D+1\right)M_{\mu\nu}\xi^{\mu}n^{\nu}$. Thence, the change of entropy~\eqref{entropy SLC} obeys
\begin{equation}
\label{DS SLC}
\Delta S_{\text{SLC}}=\Delta S_{\text{W}}+\frac{2}{D+1}\int\nabla_{\mu}\left(M^{\lambda[\mu}\alpha n^{\nu]}\xi_{\lambda}\right)n_{\nu}\text{d}^3\Sigma=\Delta S_{\text{W}}-\int M_{\mu\nu}\xi^{\mu}n^{\nu}\text{d}^3\Sigma,
\end{equation}
as required.

The non-minimal coupling makes it impossible to unambiguously split the entropy into the gravitational part (Wald entropy) and the matter part (Clausius entropy). Therefore, the term proportional to $M^{\lambda\mu}$ is added to Wald entropy precisely to account for the contributions that, for minimally coupled matter fields, would be accounted for by the Clausius entropy~\eqref{dS Clausius}. Indeed, for general relativity with minimally coupled matter, the tensor $M^{\lambda\mu}$ reads
\begin{equation}
M^{\lambda\mu}=\frac{1}{2}T^{\lambda\mu}-\frac{1}{2}\mathcal{L}_{\text{m}}g_{\mu\nu},
\end{equation}
and one can easily check that the corresponding contribution to Wald entropy precisely accounts for the Clausius entropy of the minimally coupled matter (the second term, being proportional to the metric, integrates to zero). Then, the horizon and the (in general non-minimally coupled) matter fields are treated together as one system whose entropy is given by the augmented Wald formula. While this entropy has been proposed in the context of classical gravitational dynamics, one can note the conceptual similarity with the semi-classical generalized entropy $S_{\text{gen}}$, that also contains contributions both from gravity and matter fields. The right hand side of the reversible Clausius relation, \mbox{$\Delta S_{\text{rev}}=\Delta Q/T_{\text{U}}$} in this case accounts solely for the heat flux caused by the uniformly accelerating probe moving along the stretched light cone and measuring its thermodynamics. In practice, we are free to assume that the energy of this probe is much lower than that of the system and take the limit $\Delta Q\to 0$~\cite{Mohd_2016}, leaving the condition of the vanishing reversible change of the augmented Wald entropy, $\Delta S_{\text{rev}}=0$ (again compared with the semi-classical equilibrium condition $\Delta S_{\text{gen}}=0$).

Specializing to the scalar-tensor action~\eqref{f(R) scalar} for $f\left(R\right)$ gravity, we obtain for the change of entropy over the slice of the stretched light cone
\begin{align}
\nonumber \Delta S=&\frac{1}{4G\hbar}\int_{0}^{\epsilon}\text{d}t\int_{\sigma\left(t\right)}\text{d}^{D-2}\mathcal{A}\Big[\left(\phi R_{\mu\nu}-\nabla_{\mu}\nabla_{\nu}\phi+g_{\mu\nu}\nabla^{\lambda}\nabla_{\lambda}\phi-8\pi GT_{\mu\nu}+4\pi G\mathcal{L}_{\text{m}}g_{\mu\nu}\right)\xi^{\nu} \\
&+\phi\Phi_{\;\:(\mu\nu)}^{\nu}+\nabla^{\nu}\phi\nabla_{(\nu}\xi_{\mu)}-\nabla_{\mu}\phi\nabla_{\nu}\xi^{\nu}\Big]n^{\mu}. \label{DS total}
\end{align}
As before, the first term on the second line partly corresponds to the irreversible entropy production and partly can be removed by suitable redefinitions of $\xi^{\mu}$~\cite{Parikh_2018}. The remaining terms on the second line integrate to $0$, barring the subleading $O\left(\alpha^{D-1}\right)$ curvature contributions, which can be neglected.

The first line of equation~\eqref{DS total} then accounts for the reversible change of entropy
\begin{equation}
\label{S rev}
\Delta S_{\text{rev}}=\frac{1}{4G\hbar}\int_{0}^{\epsilon}\text{d}t\int_{\sigma\left(t\right)}\text{d}^{D-2}\mathcal{A}\left(\phi R_{\mu\nu}-\nabla_{\mu}\nabla_{\nu}\phi-8\pi GT_{\mu\nu}\right)\xi^{\nu}n^{\mu}.
\end{equation}
The previously proposed prescription for the reversible part of the augmented entropy also included a term proportional to $\mathcal{L} g_{\mu\nu}$~\cite{Mohd_2016}, with $\mathcal{L}$ being the total Lagrangian. However, since $g_{\mu\nu}\xi^{\nu}n^{\mu}=0$, the contribution of $\mathcal{L} g_{\mu\nu}$ to entropy actually vanishes. This term played an important role in the original argument, making it possible to recover the effect of the scalar field potential $V\left(\phi\right)$ on the equations of motion. The derivation relied on abandoning the full expression~\eqref{S rev}, instead working just with the integrand in which the term proportional $\mathcal{L} g_{\mu\nu}$ was kept despite being irrelevant for entropy. We believe this approach to be somewhat ad hoc and instead propose a way to account for $V\left(\phi\right)$ that is more grounded in physics.

Explicitly integrating $\Delta S_{\text{rev}}$ yields
\begin{equation}
\Delta S_{\text{rev}}=\frac{\Omega_{D-2}\epsilon^2\alpha^{D-2}}{8G\hbar}E_{\mu\nu}\left(t^{\mu}t^{\nu}-\frac{1}{D}g^{\mu\nu}\right),
\end{equation}
where 
\begin{equation}
E_{\mu\nu}=\phi R_{\mu\nu}-\nabla_{\mu}\nabla_{\nu}\phi-8\pi GT_{\mu\nu},
\end{equation}
and $t^{\mu}=\partial_{t}^{\mu}$, which is arbitrary. Then, if the Clausius relation $\Delta S_{\text{rev}}=0$ applied, we would to obtain a tensorial traceless equation~\cite{Alonso:2025}
\begin{equation}
E_{\mu\nu}-\frac{1}{D}E_{\lambda}^{\ \lambda}g_{\mu\nu}=0.
\end{equation}
This equation contains no information about the scalar field potential $V\left(\phi\right)$, preventing us from recovering the full equations of motion. The solution we propose lies in treating the scalar field potential $V\left(\phi\right)$ as being imposed on the thermodynamics of the stretched light cone by its exterior. In effect, we consider the stretched light cone as an open system whose reversible entropy is not conserved, but rather proportional to the work $W$ performed on the stretched light cone by the environment
\begin{equation}
\label{S=W}
\Delta S_{\text{rev}}=W/T_{\text{U}}.
\end{equation}
As a consequence of the stretched light cone being an open system, tensor $E_{\mu\nu}$ corresponding to areal density of the entropy, but its divergence can be understood as an effective force exerted by the environment on the stretched light cone, $\nabla_{\nu}E_{\mu}^{\ \nu}=F_{\mu}$. The force corresponding to potential $V\left(\phi\right)$ is naturally proportional to its gradient. Hence we can parametrize $F_{\mu}$ as
\begin{equation}
F_{\mu}=C\nabla_{\mu}V\left(\phi\right)+F^{0}_{\mu},
\end{equation}
where $C$ is some proportionality constant and $F^{0}_{\mu}$ an undetermined additional contribution to the force. The work $W$ performed by force $F_{\mu}$ then obeys
\begin{equation}
W=\int_{0}^{\epsilon}\text{d}t\int_{\sigma\left(t\right)}\text{d}^{D-2}\mathcal{A} F_{\mu}u^{\mu}=\frac{\Omega_{D-2}}{2}\epsilon^2\alpha^{D-4}CV\left(\phi\right)+W^{0}
\end{equation}
where the first term is obtained via integration by parts, and $W^{0}$ denotes the integral of $F^{0}_{\mu}$. Plugging this result into equation~\eqref{S=W} yields
\begin{equation}
E_{\mu\nu}\left(t^{\mu}t^{\nu}-\frac{1}{D}g^{\mu\nu}\right)=\frac{8\pi G}{\alpha}CV\left(\phi\right)+\frac{16\pi G}{\Omega_{D-2}\epsilon^2\alpha^{D-3}}W^{0}.
\end{equation}
The arbitrariness of the time direction $t^{\mu}$ ensures that the following tensorial equations must hold~\cite{Alonso:2025}
\begin{equation}
\label{eom f(R) pre}
E_{\mu\nu}-\left(\frac{1}{D}E_{\lambda}^{\ \lambda}+\frac{8\pi G}{\alpha}CV\left(\phi\right)+\frac{16\pi G}{\Omega_{D-2}\epsilon^2\alpha^{D-3}}W^{0}\right)g_{\mu\nu}=0.
\end{equation}
Taking a divergence of these equations and assuming $\nabla_{\nu}T_{\mu}^{\ \nu}=0$ yields
\begin{equation}
\nabla_{\mu}\left(\frac{3}{4}\nabla^{\lambda}\nabla_{\lambda}\phi+\frac{D-2}{2D}\phi R-\frac{D-2}{2D}8\pi GT+\frac{8\pi G}{\alpha}CV\left(\phi\right)+\frac{16\pi G}{\Omega_{D-2}\epsilon^2\alpha^{D-3}}W^{0}\right)=0.
\end{equation}
Setting the work contribution $W_0$ to
\begin{equation}
W_0=-\frac{\Omega_{D-2}\epsilon^2\alpha^{D-3}}{16\pi G}\left(\frac{3}{4}\nabla^{\lambda}\nabla_{\lambda}\phi+\frac{D-2}{2D}\phi R-\frac{D-2}{2D}8\pi GT+\Lambda\right),
\end{equation}
where $\Lambda$ is an arbitrary integration constant, simplifies the divergenceless condition to
\begin{equation}
V'\left(\phi\right)\nabla_{\mu}\phi=0.
\end{equation}
This term cannot be removed by redefining $W_0$, since we defined it as the part of the work independent of the scalar field potential $V\left(\phi\right)$. Since $\nabla_{\mu}\phi$ is generically non-vanishing, we obtain an additional equation $V'\left(\phi\right)=0$ which, for the $f\left(R\right)$ choice of potential becomes
\begin{equation}
R'\left(\phi\right)\left[\phi-f'\left(R\left(\phi\right)\right)\right]=0,
\end{equation}
which is just the $f\left(R\right)$ equation of motion~\eqref{fr eom 1}. Finally, plugging $W_{0}$ into equation~\eqref{eom f(R) pre} and setting $C=\alpha/\left(16\pi G\right)$ recovers the equations of motion~\eqref{fR eom 2}
\begin{equation}
\phi R_{\mu\nu}-\nabla_{\mu}\nabla_{\nu}\phi+\left[\nabla_{\lambda}\nabla^{\lambda}\phi-\frac{1}{2}f\left(R\left(\phi\right)\right)\right]g_{\mu\nu}=8\pi GT_{\mu\nu}.
\end{equation}
The key advantage of this derivation lies in the fact that $V\left(\phi\right)$ appears organically as a source of force acting on the stretched light cone. Indeed, the relation $\nabla_{\mu}V\propto F_{\mu}$ is quite natural as is the notion that a force acting on a thermodynamic system gives rise to a work term in the first law. The proportionality constant $C$ only needs to be fixed to  recover the equations precisely in the form in which they arise from the variational principle. Forgetting the variational route to the equations of motion, one could simply absorb this constant into the definition of the potential, removing any arbitrariness from the procedure.

\paragraph{Motivation from dynamical entropy in $2$D.}

To conclude this section, we put forward a new physical motivation for the augmented Wald entropy from the recently proposed dynamical prescription~\cite{Rignon:2023,Hollands_2024,Visser_2024}. There actually exist deeper reasons to adapt this (or a similar) form of entropy for deriving gravity from thermodynamics. Since the horizon of a stretched light cone manifestly evolves in time, it may be in general inappropriate to define its Wald entropy in the same way as for stationary black holes. In particular, the stationary prescription~\eqref{SW} discards any terms proportional to the Killing vector, as they vanish on the bifurcation surface of the Killing horizon~\cite{Wald_1993,Iyer_1994}. However, the terms proportional to the boost generator $\xi^{\mu}$ integrated over the stretched light cone do not generically vanish. Including some terms of this type in the definition of Wald entropy is in fact necessary to correctly reproduce equations of motion of modified theories of gravity by thermodynamic arguments~\cite{Parikh_2018,Svesko_2019}. One then faces an ambiguity in the usual expression for Wald entropy of stationary black holes~\eqref{SW}, since it can be modified by adding a total divergence to the Lagrangian
\begin{align}
\mathcal{L}\quad&\to\quad \mathcal{L}+\sqrt{-\mathfrak{g}}\nabla_{\mu}\lambda^{\mu}, \\
\theta^{\mu}\left[\delta\right]\quad&\to\quad\theta^{\mu}\left[\delta\right]+\sqrt{-\mathfrak{g}}\delta\lambda^{\mu}, \\
Q^{\nu\mu}_{\xi}\quad&\to\quad Q^{\nu\mu}_{\xi}+2\sqrt{-\mathfrak{g}}\xi^{[\nu}\lambda^{\mu]}, \\
S_{\text{W}}\quad&\to\quad S_{\text{W}}+\frac{2\pi}{\hbar\kappa}\int_{\mathcal{H}}2\sqrt{-\mathfrak{g}}\xi^{[\nu}\lambda^{\mu]}\text{d}\sigma_{\mu\nu},
\end{align}
where $\lambda^{\mu}$ is an arbitrary vector. This shift does not affect Wald entropy of stationary black holes, since it is proportional to the Killing vector. However, it does change entropy~\eqref{SW} both for dynamical black holes and for stretched light cones (as well as for essentially any local causal horizon). It is easy to check (e.g., by following the approach discussed in subsection~\ref{Einstein from TD}) that the equations for gravitational dynamics $E_{\mu\nu}=0$ derived from thermodynamics are also affected by this shift
\begin{equation}
E_{\mu\nu}\quad\to\quad E_{\mu\nu}+\nabla_{(\nu}\lambda_{\mu)}.
\end{equation}
Of course, these equations are different from the equations of motion corresponding to the Lagrangian $\mathcal{L}+\nabla_{\mu}\lambda^{\mu}$, which are still just $E_{\mu\nu}=0$ (being unaffected by a total divergence term).

The dependence of the Noether charge on adding a total divergence to the Lagrangian disappears once we specify boundary conditions and include the contribution of the corresponding boundary Lagrangian. Then, Wald entropy becomes
\be
S_{\text{W,total}}=\frac{2\pi\hbar}{\kappa}\int_{\mathcal{H}}\left(Q_{\text{bulk}}^{\nu\mu}-j^{\nu\mu}_{\text{bd}}\right)\text{d}\sigma_{\mu\nu},
\ee
where $j^{\nu\mu}_{\text{bd}}$ denotes the boundary Noether current 
\be
j^{\nu\mu}_{\text{bd}}=\theta^{\nu\mu}_{\text{bd}}\left[\pounds_{\xi}\right]-L_{\text{bd}}\sqrt{\vert\mathfrak{h}\vert}\xi^{[\nu}n^{\mu]},
\ee
where $\theta^{\nu\mu}_{\text{bd}}$ is the boundary symplectic potential, $L_{\text{bd}}$ the boundary Lagrangian, $\mathfrak{h}$ the determinant of the boundary metric and $n^{\nu}$ the unit, outer-pointing normal to the boundary. We leave the details of this construction for appendix~\ref{Wald entropy review}.

For simplicity, and since this is the case relevant for the next section, we discuss the recovery of augmented entropy in $2$D scalar-tensor gravity with a kinetic term for the scalar field\footnote{One may notice that this expression is equivalent to Polyakov action studied in the following section}
\be
I_{\text{2D,bulk}}=\frac{1}{16\pi G}\int\left(\phi R+\nabla_{\lambda}\phi\nabla^{\lambda}\phi\right)\sqrt{-\mathfrak{g}}\text{d}x^2.
\ee
For any standard boundary conditions for the metric (Dirichlet, Neumann or conformal), the boundary Lagrangian is proportional to the scalar extrinsic curvature. It is easy to check that the extrinsic curvature of the stretched light cone vanishes in $2$D, $\nabla_{\mu}n^{\mu}=0$. Consequently, the choice of the boundary conditions for the metric is irrelevant for stretched light cone entropy in $2$D and we do not need to discuss it further. For the scalar field, we choose Neumann boundary conditions, $\delta\left(n^{\lambda}\nabla_{\lambda}\phi\right)=0$. This choice is natural, as it fixes the boundary projection of the contribution of the kinetic term $\nabla_{\lambda}\phi\nabla^{\lambda}\phi$ to the equations of motion, whose contribution to Wald entropy we are trying to recover. The corresponding action on a timelike boundary $\Sigma$ (in our case on the stretched light cone) reads
\be
I_{\text{2D,bd}}=-\frac{1}{8\pi G}\int_{\Sigma}\phi n^{\lambda}\nabla_{\lambda}\phi\text{d}\Sigma.
\ee
The corresponding boundary symplectic potential $\theta^{\nu\mu}_{\text{bd}}$ vanishes. Wald entropy of the stretched light cone then reads
\be
\label{S_W bd}
S_{\text{W,total}}=\frac{\alpha}{4G\hbar}\left(\phi\nabla^{\nu}\xi^{\mu}-2\nabla^{\nu}\phi\xi^{\mu}-\phi n^{\lambda}\nabla_{\lambda}\phi\xi^{[\nu}n^{\mu]}\right)u_{[\mu}n_{\nu]}.
\ee
The change of entropy between times $t=0$ and $t=\epsilon$ equals (we use that $\nabla_{(\mu}\xi_{\nu)}=0$ in $2$D)
\be
\Delta S_{\text{W,total}}=\frac{\alpha}{4G\hbar}\int_{0}^{\epsilon}\text{d}t\left[n^{\mu}\xi^{\nu}\left(\nabla_{\mu}\nabla_{\nu}\phi-\nabla_{\mu}\phi\nabla_{\nu}\phi\right)-\phi\pounds_{\xi}\left(n^{\lambda}\nabla_{\lambda}\phi\right)\right].
\ee
The last term vanishes as a consequence of the Neumann boundary condition we imposed, $\delta\left(n^{\lambda}\nabla_{\lambda}\phi\right)=0$, when we set $\delta=\pounds_{\xi}$. Then, the change of entropy simplifies to
\be
\Delta S_{\text{W,total}}=\frac{\alpha}{4G\hbar}\int_{0}^{\epsilon}\text{d}tn^{\mu}\xi^{\nu}\left(\nabla_{\mu}\nabla_{\nu}\phi-\nabla_{\mu}\phi\nabla_{\nu}\phi\right).
\ee

Let us now compute tensor $M^{\lambda\mu}$ appearing in the augmented Wald entropy~\eqref{S aug} for null surfaces. It reads
\begin{equation}
M^{\mu\nu}=\frac{1}{16\pi G}\nabla^{\mu}\phi\nabla^{\nu}\phi.
\end{equation}
The total change of augmented entropy equals (see equation~\eqref{S aug})
\begin{equation}
\Delta S=\frac{\alpha}{4G\hbar}\int_{0}^{\epsilon}\text{d}tn^{\mu}\xi^{\nu}\left(\nabla_{\mu}\nabla_{\nu}\phi-\nabla_{\mu}\phi\nabla_{\nu}\phi\right),
\end{equation}
in precise agreement with $\Delta S_{\text{W,total}}$. Thence, for scalar-tensor gravity in $2$D, augmented entropy is nothing but Wald entropy with Neumann boundary conditions imposed on the scalar field. Of course, the analysis we performed here for one special example can serve only as a motivation for equating augmented and Wald entropies, and a much more careful study is still necessary. Nevertheless, for our purposes here, the computation we performed covers all the cases we study in the next section (the key point is just the behavior of the kinetic term for the scalar field).

In higher dimensions, imposing Neumann boundary conditions on the scalar field straightforwardly yields the kinetic term contribution to Wald entropy. However, a full analysis of the metric boundary conditions becomes more involved. We plan it report on it in a future work.

\section{Semi-classical 2D gravity from thermodynamics} \label{sec:semi2Dgrav}

So far, we have discussed thermodynamics of non-minimally coupled gravity in the fully classical context. The algorithm we outlined for deriving the equations of motion of scalar-tensor theories can be readily applied in the semi-classical setting, to explore the problem of the backreaction of quantum field on the classical gravitational spacetime. While the problem becomes too involved in $4D$ (as we discuss in the next section), the $2$D case turns out to be fully treatable and exposes some interesting features valid also for general higher-dimensional cases. Upon reviewing the backreaction problem in $2$D in subsection~\ref{2D gravity}, we proceed to derive the corresponding equations of motion from the generalized entropy in subsection~\ref{derivation}.

\subsection{Semi-classical 2D gravity}
\label{2D gravity}

Scalar-tensor $2$D JT gravity is an interesting arena for studying the semiclassical backreaction problem for several reasons. First, the entire backreaction of the quantum fields is captured by the conformal anomaly and can be encoded in the local Polyakov action~\cite{Polyakov:1981,Almheiri:2015}. Second, the resulting semiclassical theory is stable and has a clearly defined regime of validity~\cite{Pedraza:2021cvx}. Third, it allows analytical black hole solutions that can serve as a toy model providing hints for the behavior of semiclassical black holes in higher dimensions, e.g. in regards to the information recovery~\cite{Russo:1992,Almheiri:2020,Pedraza:2021cvx}.

Thermodynamics of JT gravity also has remarkable properties. In the semiclassical context, the relevant thermodynamic quantity is the generalized entropy $S_{\text{gen}}$ which combines both the entropy of the classical gravitational field and of the exterior quantum fields. The semiclassical equations of motion should then follow from the equilibrium condition $\Delta S_{\text{gen,rev}}=0$, i.e., the reversible change of the generalized entropy vanishes. However, explicitly computing the generalized entropy represents a formidable task. While the induced gravity paradigm identifies the generalized entropy with the total entanglement entropy $S_{\text{EE}}$ of the spacetime~\cite{Susskind:1994,Jacobson_1994}, an explicit evaluation of $S_{\text{EE}}$ remains equally challenging (due to its nonlocal nature for nonconformal quantum fields~\cite{Jacobson_2016,Speranza:2016,Casini:2016} and to the ambiguities in the quantum expectation value of the stress-energy tensor~\cite{Wald:1994}). The situation drastically simplifies in $2$D. It has been shown that Wald entropy for black hole solutions of semiclassical JT gravity coincides with their total entanglement entropy, even in time-dependent situations~\cite{Pedraza:2021cvx}. Then, Wald entropy also agrees with the generalized entropy. For the case of locally constructed stretched light cones, the generalized entropy turns out to be given by the augmented Wald entropy expression~\eqref{S aug}, as we argue in the following. The augmented Wald entropy then suffices to recover the complete semiclassical equations of motion fully capturing the bakreaction of quantum fields. We stress that, while we compute the total entanglement entropy via Wald prescription~\eqref{S aug}, its form can in fact by supported by purely kinematic considerations, without any input from gravitational dynamics. Indeed, classical $2$D JT gravity can be interpreted as a low-energy model for the classical dynamics of a class of higher-dimensional charged near-extremal black holes and branes~\cite{Achucarro:1993,Fabbri:1995,Nayak:2018,Sachdev:2019}. In this case, the entanglement entropy of the higher dimensional black hole corresponds to Wald entropy of JT gravity. Since the entanglement entropy scaling with area follows from quantum field theory in the presence of a horizon, the form of Wald entropy of JT gravity has a motivation completely independent of gravitational dynamics. The form of the conformal anomaly contribution to entanglement entropy is likewise a generic prediction for a conformal quantum field theory on a curved background, obtained without any reference to gravitational dynamics~\cite{Duff:1994,Pedraza:2021cvx}. Therefore, our results in this and the subsequent section follow just from kinematics of conformal quantum fields in a curved background. The only other necessary ingredient is the Einstein equivalence principle which justifies approximating the local vacuum state at any regular spacetime point by Minkowski vacuum~\cite{Chirco_2010,Alonso:2024}.

Let us now recall the basics of semiclassical JT gravity. The classical JT action is given by
\begin{equation}
\label{eq:jt}
I_\text{dil}=\frac{1}{16\pi G}\int_\mathcal{M}\sqrt{-\mathfrak{g}}\text{d}^{2}x\left[ \phi_0 R+\phi\left(R+\frac{2}{L^2}\right)+\int_\mathcal{M}\sqrt{-\mathfrak{g}}\text{d}^{2}x \mathcal{L}_{\text{m}} \right],
\end{equation}
where $\phi$ is a classical dilaton field, $\phi_0$ a constant\footnote{If we understand classical $2$D JT gravity as a low-energy model for higher-dimensional charged near-extremal black holes and branes~\cite{Achucarro:1993,Fabbri:1995,Nayak:2018,Sachdev:2019}, $\phi_0$ is proportional to the entropy of this higher-dimensional extremal black hole. While our results are not directly related to the higher-dimensional interpretation of JT gravity, we keep $\phi_0$ explicit to show that the thermodynamic derivation also works in this case.}, $\Lambda=-1/L^2$ is the (negative) cosmological constant, $\sqrt{-\mathfrak{g}}\text{d}^{2}x$ denotes the volume $2$-form, and $\mathcal{L}_{\text{m}}$ is the matter Lagrangian. 

The effective action $I_{\text{eff}}$ represents a useful way to describe coupling of a quantum and classical system, i.e., precisely the situation one faces in semi-classical gravity. It can be defined in the context of Euclidean path integral formalism, schematically, as
\begin{equation}
e^{-I_{\text{eff}}\left[J\right]}=\int\mathcal{D}\chi e^{-I_{\text{cl}}\left[\chi,J\right]},
\end{equation}
where $I_{\text{cl}}$ is the classical Euclidean action, $\chi$ the quantum fields, $J$ the classical fields, and $\mathcal{D}\chi$ a suitably defined path integral measure. The effective action then depends only on the classical fields $J$ and captures the backreaction of quantum fields $\chi$ on them. In this way, the one-loop semiclassical corrections to JT gravity induced by a conformal field with a central charge $c$ are fully characterised by the local Polyakov effective action of the form
\begin{equation}
\label{eq:poly}
I_\text{Poly}=-\frac{c}{24\pi} \int_{\mathcal{M}}\sqrt{-\mathfrak{g}}\text{d}^2x\left[\chi R+(\nabla\chi)^2-\frac{\lambda}{L^2}\right],
\end{equation}
where $\lambda$ is an arbitrary constant which corresponds to an ambiguity present in the trace anomaly. The full equations of motion then read
\begin{align}
\begin{split}
\label{eq:eom}
E^{\mu\nu}_{g}=&\left(\nabla^\nu\nabla^\mu\phi-\Box \phi g^{\mu\nu}+\frac{1}{L^2}\phi\right)\\&+8\pi G\frac{c}{12\pi}\left[\nabla^\mu\chi\nabla^\nu\chi-\frac{1}{2}(\nabla\chi)^2g^{\mu\nu}-\nabla^\nu \nabla^\mu\chi+g^{\mu\nu}\Box\chi+\frac{\lambda}{2L^2}g^{\mu\nu} \right]+8\pi G\langle T^{\mu\nu}\rangle=0, \\
E_\phi =&\frac{1}{16\pi G}\left(R+\frac{2}{L^2}\right)=0, \\
E_\chi =&\frac{c}{24\pi}(\frac{1}{2}R-\Box \chi)=0.
\end{split}
\end{align}
On shell, the trace of the stress-energy tensor of the field $\chi$
\begin{equation}
T^{\chi}_{\mu\nu}=\frac{c}{12\pi}\left[\nabla^\mu\chi\nabla^\nu\chi-\frac{1}{2}(\nabla\chi)^2g^{\mu\nu}-\nabla^\nu \nabla^\mu\chi+g^{\mu\nu}\Box\chi+\frac{\lambda}{L^2}g^{\mu\nu}\right],
\end{equation}
reproduces the trace anomaly contribution of $c$ conformal fields in $2$D, i.e.,
\begin{equation}
\label{2D anomaly}
g^{\mu\nu}T^{\chi}_{\mu\nu}\big\vert_{\text{on shell}}=\frac{c}{24\pi}\left(-\frac{2}{L^2}+\frac{2\lambda}{L^2}\right).
\end{equation}

Before proceeding, we need to address the regime of validity of this semiclassical action. The central charge $c$ must be much greater than $1$ so that the loop corrections associated with $\chi$ can be disregarded~\cite{Almheiri:2015}. At the same time, $c$ must be sufficiently small to still represent a correction to the classical dynamics rather than a dominant effect. The precise meaning of this statement is context-dependent (see~\cite{Pedraza:2021cvx} for the discussion of the black hole case). While the limit of very large $c$ is also interesting to study, it would break the self-consistency of the backreaction approach we adopt. Nevertheless, braneworld solutions with very large $c$ appear to still possess the generalized entropy given by the semiclassical expression we use~\cite{Panella:2023}, suggesting that our results might be applicable even in this regime.

\subsection{Semiclassical equations of motion from thermodynamics}
\label{derivation}

Upon recalling the relevant features of the semiclassical JT gravity, we proceed to show that its equations of motion are fully encoded in the local equilibrium condition $\Delta S_{\text{gen}}=0$. We apply the method developed in section~\ref{classical}. Based on the equivalence of generalized and Wald entropies for $2$D black holes~\cite{Pedraza:2021cvx}, we compute the conformal anomaly contribution to $S_{\text{gen}}$ as Wald entropy with Neumann boundary condition imposed on the Polyakov scalar field. We return to the appropriateness of this identification at the end of the subsection.

We work in the framework of the local thermodynamics of stretched light cones reviewed in subsection~\ref{Einstein from TD}. In $2$D, geometry of stretched light cones is similar, although some important differences occur. Consider a point $P$ in the manifold $\mathcal{M}$ and Riemann normal neighborhood around it. The metric tensor admits an expansion in local Riemann normal coordinates $x=(t,r)$ as
\begin{equation}
g_{\mu\nu}(x)=\eta_{\mu\nu}-\frac{1}{3}R_{\mu\nu\rho\sigma}(P)x^{\rho}x^{\sigma}+\mathcal{O}(x^3).
\end{equation}
As before, we define the future-oriented light cone as a $1$-dimensional hypersurface given by the condition $r^2-t^2=0$, i.e. where the boost generator obeys
\begin{equation}
\xi=r\partial_t+t\partial_r,
\end{equation}
with $\{\partial_t,\partial_r\}$ being the basic of $T_p\mathcal{M}$. Note that in a higher-dimensional case this generator corresponds to a spherical boost and as such it is not an isometry of the Minkowski space. However, in the case of two-dimensional gravity spherical and Cartesian directions are the same (since the constant $t, r$ surfaces reduce to points). The field $\xi$ is thus an approximate Killing vector and satisfies, up to higher orders in $x$, the Killing equation and identity
\begin{align}
\label{Killing eq}
&\nabla_\mu\xi_\nu+\nabla_\nu\xi_\mu=O(x^2),\\
\label{Killing id}
&\nabla_\mu\nabla_\nu\xi_\sigma=R\indices{^\alpha_{\mu\nu\sigma}}\xi_\alpha +O(x).
\end{align}
These equations would not hold in higher dimensions, although there exist similar approximations for truly spherical boosts, reviewed in subsection~\ref{Einstein from TD} (see also~\cite{Parikh_2018}).

To construct our stretched light cone, we now fix a length scale $\alpha$ such that it is much smaller than curvature at $P$. We consider a class of timelike curves with a tangent vector proportional to $\xi$ and choose one with acceleration $a=1/\alpha$ at $t=0$. Finally, we fix a timescale $\epsilon\ll\alpha$ so that for $t\in(0,\epsilon)$ the acceleration is approximately constant. The congruence of these curves forms a one-dimensional hypersurface which is, up to curvature-dependent deformations, the hyperboloid
\begin{equation}
r^2-t^2=\alpha^2.
\end{equation}
Since the acceleration is approximately constant, this surface has a well-defined temperature at the time interval $0<t<\epsilon$.

We take the leading order entanglement entropy of the stretched light cone to be proportional to the scalar field $S_0=\eta\phi$. This choice of entropy is motivated by expressing entanglement entropy associated with a horizon of a higher-dimensional extremal black hole. The evolution of the horizon can be described as an effective AdS$2$ geometry. The entanglement entropy of the black hole translates to this geometry precisely as $S_0=\eta\phi$, where $\phi$ appears as a scale factor~\cite{Achucarro:1993,Fabbri:1995,Nayak:2018,Sachdev:2019}. Since $2$D entanglement entropy is proportional to the value of a scalar field in this special situation, it becomes natural to expect it takes this form in general. We further define the $2$D gravitational constant $G$ in terms of the proportionality constant, i.e., $G=1/\left(4\hbar\eta\right)$, obtaining
\be
S_0=\frac{\phi+\phi_0}{4G\hbar},
\ee
where, for the sake of easier comparison with literature on JT gravity, we split the scalar field into a constant part $\phi_0$ corresponding to the potential higher dimensional black hole and variable part $\phi$.

The second contribution to the total generalized entropy comes from quantum matter present in the $2$D spacetime. For simplicity, we assume that it corresponds to conformally invariant quantum fields. Then, the change of von Neumann entropy along the evolution of the stretched light cone equals (see equation~\eqref{S vN})
\be
\Delta S_{\text{vN}}=\frac{2\pi\alpha}{\hbar}\int\langle T_{\mu\nu}\rangle\xi^{\nu}\text{d}\Sigma^{\mu},
\ee
where $\langle T_{\mu\nu}\rangle$ represents the expectation value of the stress-energy operator defined with respect to the local Minkowski vacuum state.

The von Neumann entropy of conformal fields further acquires a contribution from the conformal anomaly. The total local matter Lagrangian capturing the conformal anomaly reads
\begin{equation}
\label{2D L}
\mathcal{L}_{\text{2D}}=\mathcal{L}_{\text{m}}-\frac{c}{24\pi}\left[\chi R+(\nabla\chi)^2-\frac{\lambda}{L^2}\right],
\end{equation}
where $\chi$ is the Polyakov field. Inspired by the equivalence of entanglement and Wald entropy of semiclassical black holes in $2$D~\cite{Pedraza:2021cvx}, we compute the Polyakov entropy contribution via Wald prescription. Since the stretched light cone is only quasi-stationary, even near the bifurcation suurface at $t=0$, the boundary Lagrangian contribution to Wald entropy must be taken into account to obtain an unambiguous result. Inspired by our analysis at the end of the previous section, we choose Neumann boundary condition for the Polyakov field, i.e., $\delta\left(n^{\mu}\nabla_{\mu}\chi\right)=0$. The resulting Wald entropy has the same form as equation~\eqref{S_W bd} with the scalar field being $\chi$ and the coupling constant equal to $-c/\left(24\pi\right)$ rather than $1/\left(16\pi G\right)$, i.e.,
\begin{equation}
\label{S_P}
S_{\text{P}}=-\frac{c\alpha}{6}\left(\chi\nabla^{\nu}\xi^{\mu}-2\nabla^{\nu}\chi\xi^{\mu}+\nabla^{\mu}\chi\nabla^{\nu}\chi\right)u_{[\mu}n_{\nu]}.
\end{equation}
We now have the complete generalized entropy prescription.

The change of the total generalized entropy $S=S_0+S_{\text{vN}}+S_{\text{P}}$ along the stretched light cone between times $t=0$ and $t=\epsilon$ reads, using the generalized Stokes theorem,
\begin{equation}
\Delta S=\int (\nabla_\mu s^{\mu\nu}- M^{\mu\nu}\xi_\mu) n_\nu\dd\Sigma,
\end{equation}
where the integral is performed over a segment of the $1$-dimensional surface of the stretched light cone bounded by times $t=0$ and $t=\epsilon$. For $\nabla_\mu s^{\mu\nu}$ it holds
\begin{equation}
\nabla_\mu s^{\mu\nu}=P^{\nu\mu\rho\sigma}(R\indices{^\alpha_{\mu\rho\sigma}}\xi_\alpha+\Phi_{\mu\rho\sigma})-2\nabla_{(\mu}\xi_{\rho)}\nabla_\sigma P^{\nu\mu\sigma\rho}-2\xi_\rho\nabla_\mu\nabla_\sigma P^{\nu\mu\sigma\rho},
\end{equation}
where 
\be
M^{\mu\nu}=\frac{c}{24\pi}\nabla^\mu\chi\nabla^\nu\chi+\frac{1}{2}\langle T^{\mu\nu}\rangle,
\ee
and $P^{\nu\mu\rho\sigma}$ reads
\be
\label{P JT}
P^{\nu\mu\rho\sigma}=\left[\frac{\alpha}{8G\hbar}\left(\phi_0+\phi\right)-\frac{c\alpha}{12}\chi\right]\left(g^{\mu\sigma}g^{\nu\rho}-g^{\mu\rho}g^{\nu\sigma}\right),
\ee
and the first term is obtained using the corrected Killing identity. As shown in subsection~\ref{Einstein from TD} the integral of the expression $\Phi_{\mu\sigma\rho}$ is proportional to $(D-2)$ (see equation~\eqref{irr}). Since our theory is in $2$D, this contribution is exactly $0$ and we may drop this term. A geometrical interpretation of this fact is that in one spatial dimension, there is no difference between ``Cartesian" and spherical boosts and $\xi^{\mu}$ is an exact Killing vector obeying the Killing identity. Therefore, $\Phi_{\mu\sigma\rho}$ actually identically vanishes. Similarly, the term $-2\nabla_{(\mu}\xi_{\rho)}\nabla_\sigma P^{\nu\mu\sigma\rho}$ vanishes as a direct consequence of the Killing equation, which is exactly satisfied in $2$D.

In total, we have for the change of the generalized entropy
\begin{equation}
\Delta S=\int\left(P^{\nu\lambda\rho\sigma}R\indices{^\mu_{\lambda\rho\sigma}}-2\nabla_\rho\nabla_\sigma P^{\nu\rho\sigma\mu}+\frac{2cG}{3}\nabla^\mu\chi\nabla^\nu\chi-\frac{1}{2}\langle T^{\mu\nu}\rangle\right)\xi_\mu n_\nu\dd\Sigma.
\end{equation}
The absence of irreversible entropy production means that the Clausius relation requires $\Delta S=W/T_{\text{U}}$. As in section~\ref{classical}, the work term consist of a part proportional to the scalar field potential $V\left(\phi\right)$ and some remainder $W^0$. Then, proceeding exactly as in section~\ref{classical}, we integrate the entropy prescription and consider that our change of time direction is arbitrary to convert the result into a tensorial equation. We obtain
\begin{align}
\nonumber &P^{\nu\lambda\rho\sigma}R^{\mu}_{\lambda\rho\sigma}-2\nabla_{\rho}\nabla_{\sigma} P^{\nu\rho\sigma\mu}+\frac{2cG}{3}\nabla^{\mu}\chi\nabla^{\nu}\chi-\frac{1}{2}\langle T_{\mu\nu}\rangle-\frac{1}{2}g_{\mu\nu}\bigg(P_{\alpha}^{\ \lambda\rho\sigma}R^{\alpha}_{\lambda\rho\sigma}-2\nabla_\rho\nabla_\sigma P_{\lambda}^{\ \rho\sigma\lambda} \\
&+\frac{2cG}{3}\nabla_\lambda\chi\nabla^\lambda\chi-\frac{1}{2}\langle T_{\lambda}^{\ \lambda}\rangle\bigg)+\left(\frac{16\pi G\alpha}{\epsilon^2}W^0+\frac{1}{2}V\left(\phi\right)\right)g^{\mu\nu}=0. \label{eq:claus}
\end{align}

The expression~\eqref{P JT} for $P^{\nu\lambda\rho\sigma}$ leads to
\begin{align}
&\nabla^\alpha\nabla^\beta P_{\mu\alpha\beta\nu}=-\frac{1}{2}(\nabla_\nu\nabla_\mu-g_{\mu\nu}\Box)\left(\phi-\frac{2cG}{3}\chi\right),\\
&P\indices{_\mu^{\alpha\beta\gamma}}R_{\nu\alpha\beta\gamma}=R_{\mu\nu}\left(\phi_0+\phi+\frac{2cG}{3}\chi\right)=\frac{1}{2}Rg_{\mu\nu}\left(\phi_0+\phi-\frac{2cG}{3}\chi\right).
\end{align}
Eq.~(\ref{eq:claus}) then becomes
\begin{align}
\label{eq:main_eq}
\nonumber &\left(\nabla_\mu\nabla_\nu-\frac{1}{2}g_{\mu\nu}\Box\right)\left(\phi-\frac{2cG}{3}\chi\right)+\frac{2cG}{3}\nabla^\mu\chi\nabla^\nu\chi-8\pi G\langle T_{\mu\nu}\rangle \\
&+\left[4\pi G\langle T_{\lambda}^{\ \lambda}\rangle-\frac{cG}{3}(\nabla\chi)^2+V\left(\phi\right)+\frac{16\pi G\alpha}{\epsilon^2}W^0\right]g_{\mu\nu}=0.
\end{align}
Taking the covariant divergence and imposing conservation of the stress-energy tensor results in
\begin{align}
\nonumber &\frac{2cG}{3}\nabla^\mu\chi\left(\frac{1}{2}R-\Box\chi\right)+\nabla_{\mu}\phi\left[\frac{1}{2}R-V'\left(\phi\right)\right] \\
&+\nabla_{\mu}\left[\frac{16\pi G\alpha}{\epsilon^2}W^0+4\pi G\langle T_{\lambda}^{\ \lambda}\rangle+\frac{1}{2}\Box\left(\phi-\frac{2cG}{3}\chi\right)\right]=0.
\end{align}
Satisfying this constraint while keeping the fields $\phi$ and $\chi$ independent yields two conditions
\begin{align}
W^0&=-\frac{\epsilon^2}{4\alpha}\langle T_{\lambda}^{\ \lambda}\rangle-\frac{\epsilon^2}{32\pi G\alpha}\Box\left(\phi-\frac{2cG}{3}\chi\right), \\
\frac{1}{2}R-V'\left(\phi\right)&=0, \\
\frac{1}{2}R-\Box\chi&=0.
\end{align}
The last two conditions are the equations of motion corresponding to variations with respect to $\phi$ and $\chi$. Plugging $W^0$ back into the tensorial equations~\eqref{eq:main_eq} yields
\begin{align}
\nonumber &\left(\nabla_\mu\nabla_\nu-g_{\mu\nu}\Box\right)\left(\phi-\frac{2cG}{3}\chi\right)+\frac{2cG}{3}\nabla^\mu\chi\nabla^\nu\chi-8\pi G\langle T_{\mu\nu}\rangle \\
&+\left(-\frac{cG}{3}(\nabla\chi)^2+V\left(\phi\right)\right)g_{\mu\nu}=0.
\end{align}
We recognize these equations as the equations of motion of semiclassical JT gravity with arbitrary dilaton potential $V\left(\phi\right)$. We obtain the result for standard JT gravity simply by setting $V\left(\phi\right)=\phi/L^2$. On shell, the trace of the equations of motion then correctly recovers the conformal anomaly~\eqref{2D anomaly}.

We recovered the semiclassical dynamics of JT gravity fully accounting for the backreaction of the quantum fields augmented Wald entropy $S_{\text{P}}$~\eqref{S_P}. The appropriate equilibrium condition in $2$D should involve the generalized entropy $S_{\text{gen}}$, i.e, $\Delta S_{\text{gen}}=W/T_{\text{U}}$. Presumably, this condition then allows the recovery of the full semiclassical dynamics. Thence, we have shown the equivalence between augmented Wald entropy and the generalized entropy in $2$D, $S_{\text{SLC}}=S_{\text{gen}}$. As we shown in subsection~\ref{subsec:n-m-g} the augmented Wald entropy in $2$D is further equal to the dynamical Wald entropy with Neumann boundary conditions imposed on the scalar field. The generalized entropy in $2$D is then simply the semiclassical dynamical Wald entropy. This observation generalizes the previous result that the generalized entropy of semiclassical black holes in JT gravity corresponds to their Wald entropy. As we discuss in the next section, this straightforward identification does not appear to carry over to higher dimensions.

\section{Semi-classical Einstein gravity?}
\label{4D}

We have developed a way to study semi-classical gravitational dynamics in two dimensions via thermodynamic methods. Our approach was made possible by the simple structure of the conformal anomaly in $2$D. The situation in $4D$ becomes far more complicated due to the presence of the curvature squared contributions to the conformal anomaly~\cite{Riegert:1984,Fradkin:1984,Duff:1994}. These terms cannot be obtained from a local, metric, diffeomorphism-invariant action. Nevertheless, local scalar-tensor prescriptions for the effective action reproducing the conformal anomaly in $4D$ have been put forward~\cite{Riegert:1984,Gabadadze:2023,Lowe:2025}, although they cannot account for the $R^2$ contribution to the conformal anomaly~\cite{Riegert:1984}. The physical implications of the conformal anomaly were then studied in the context of black hole physics~\cite{Cai:2010,Calmet:2017,Calmet:2020,Fernandes:2023,Ho:2024,Navarro:2024}, as well as in cosmology~\cite{Calmet:2024}. In $2$D, the effective action reproducing the conformal anomaly fully encodes the semi-classical backreaction~\cite{Pedraza:2021cvx}. However, in $4D$ there appears to be a mismatch between the effective dynamics and direct calculations of the renormalised stress-energy tensor~\cite{Arrechea:2024}. Moreover, the effective dynamics fails to reproduce the physically expected behaviour of the Unruh vacuum state in the asymptotic infinity~\cite{Bardeen:2018,Lowe:2025}. These discrepancies suggest that the effective action no longer fully captures the semi-classical backreaction. Nevertheless, it provides valuable insights into the semi-classical physics of black holes.

Herein, we discuss reproducing the equations of motion which give rise to the conformal anomaly from thermodynamics. The effects of the conformal anomaly on black hole entropy have been discussed in the literature~\cite{Aros:2010,Aros:2013,Majhi:2014}, and we easily construct the expression for the augmented Wald entropy (defined by equation~\eqref{DS SLC}). The open question is whether the resulting entropy corresponds to the generalized entropy in semi-classical Einstein gravity. In the $2$D case, the equivalence of both entropies has been shown explicitly~\cite{Pedraza:2021cvx}. In $4D$, the question is significantly more involved and no explicit comparison between entropies has been performed as of yet. Thermodynamic recovery of the equations of motion actually offers a partial answer. On the one hand, our ability to reconstruct the equations from the augmented Wald entropy shows that it is in one-to-one correspondence with the effective dynamics. On the other hand, we mentioned that some results in black hole physics suggest that the effective dynamics does not fully capture the semi-classical dynamics of general relativity~\cite{Arrechea:2024}. Conversely, we expect that the complete expression for generalized entropy should differ from augmented Wald entropy, to be equivalent to the full semi-classical dynamics.

In four spacetime dimensions, the conformal anomaly obeys~\cite{Duff:1994}
\begin{equation}
\label{anomaly 4D}
\langle T_{\mu}^{\;\:\mu}\rangle = aC_{\mu\nu\rho\sigma}C^{\mu\nu\rho\sigma}+b\mathcal{G}-c\Box R+dR^2,
\end{equation}
with $\mathcal{G}$ being the Gauss-Bonnet term
\begin{equation}
\mathcal{G}=R^2-4R_{\lambda\rho}R^{\lambda\rho}+R_{\lambda\rho\sigma\tau}R^{\lambda\rho\sigma\tau}.
\end{equation}
Coefficients $a$, $b$ are determined by the number of massless scalar bosons $N_{\text{s}}$, massless, spin $1/2$ fermions $N_{\text{f}}$ and massless vector bosons $N_{\text{v}}$, i.e.,
\begin{align}
a=&\frac{1}{1920\pi^2}\left(N_{\text{s}}+6N_{\text{f}}+12N_{\text{v}}\right), \\
b=&-\frac{1}{5760\pi^2}\left(N_{\text{s}}+11N_{\text{f}}+62N_{\text{v}}\right).
\end{align}
The coefficient $c$ depends on the regularisation scheme. Lastly, the coefficient $d$ is not sourced by fields with a spin less than or equal to $1$~\cite{Duff:1994}, and, following the previous studies~\cite{Riegert:1984,Lowe:2025}, we set it zero. However, if the term $dR^2$ does appear, it cannot be recovered from any diffeomorphism-invariant action, neither local nor nonlocal~\cite{Riegert:1984}. This issue represents one of the significant drawbacks of the conformal anomaly approach to effective semi-classical dynamics.

For $d=0$, there exists a local effective Lagrangian that correctly reproduces the conformal anomaly contribution to the gravitational equations of motion~\cite{Lowe:2025}. It reads\footnote{We have applied integration by parts to rewrite the term $\phi\Box R$ into $R\Box\phi$ (equivalent up to total derivative terms), for which it is easier to compute the corresponding Noether charge.}
\begin{align}
\nonumber \mathcal{L}_{\text{GR+A}}=&\frac{1}{16\pi}\left(R-2\Lambda\right)+\frac{3c-2b}{576\pi^2}R^2+bR^{\lambda\rho}\nabla_{\lambda}\phi\nabla_{\rho}\phi-\frac{b}{3}R\nabla_{\lambda}\phi\nabla^{\lambda}\phi-\frac{b}{2}\Box\phi\Box\phi \\
&+\frac{\phi}{8\pi}\left[\left(a+b\right)R_{\lambda\rho\sigma\tau}R^{\lambda\rho\sigma\tau}-2\left(a+2b\right)R_{\lambda\rho}R^{\lambda\rho}+\left(\frac{1}{3}a+b\right)R^2\right]-\frac{b}{12\pi}R\Box\phi. \label{anomaly 4D action}
\end{align}
The absence of the kinetic term for the scalar field indicates that the theory is strongly coupled at arbitrarily low energy scales, implying instability of the Minkowski vacuum~\cite{Gabadadze:2023}. Nevertheless, the action is of interest as a toy model for the semi-classical gravitational dynamics in $4D$. The corresponding Noether charge tensor reads
\begin{equation}
Q^{\nu\mu}_{\xi,\text{GR+A}}=-2P^{\mu\nu\rho\sigma}\nabla_{\rho}\xi_{\sigma}+4\nabla_{\rho}P^{\mu\nu\rho\sigma}\xi_{\sigma}+b\nabla^{[\nu}\phi\xi^{\mu]}\left(\frac{1}{6\pi}R+2\Box\phi\right),
\end{equation}
where
\begin{align}
\nonumber P^{\mu\nu\rho\sigma}=&\frac{\partial \mathcal{L}_{\text{GR+A}}}{\partial R_{\mu\nu\rho\sigma}} \\
\nonumber =&\left(\frac{1}{16\pi}+\frac{a+3b}{12\pi}\phi R-\frac{b}{3}\nabla_{\lambda}\phi\nabla^{\lambda}\phi-\frac{b\Box\phi}{12\pi}+\frac{3c-2b}{288\pi^2}R\right)g^{\rho[\mu}g^{\nu]\sigma}+\frac{\left(a+b\right)\phi}{4\pi}R^{\mu\nu\rho\sigma} \\
& +\frac{\left(a+2b\right)\phi}{4\pi}\left(R^{\rho[\nu}g^{\mu]\sigma}+R^{\sigma[\mu}g^{\nu]\rho}\right)+\frac{b}{2}\left(\nabla^{\mu}\phi g^{\nu[\sigma}\nabla^{\rho]}\phi-\nabla^{\nu}\phi g^{\mu[\sigma}\nabla^{\rho]}\phi\right).
\end{align}
The contribution $b\nabla^{[\nu}\phi\xi^{\mu]}\left(\frac{1}{6\pi}R+2\Box\phi\right)$ not expressible in terms of $P^{\mu\nu\rho\sigma}$ comes from the variations of $\Box\phi$.

Aside from the Noether charge contribution, the augmented Wald entropy includes a term accounting for the non-minimally coupled kinetic term of the scalar field, $R\nabla_{\lambda}\phi\nabla^{\lambda}\phi$. The total augmented entropy then reads
\begin{equation}
S_{\text{SLC}}=\frac{2\pi}{\hbar\kappa}\int Q^{\nu\mu}_{\xi,\text{GR+A}}\text{d}\sigma_{\mu\nu}+\frac{2}{D+1}\int M^{\lambda[\mu}\alpha n^{\nu]}\xi_{\lambda}\text{d}\sigma_{\mu\nu},
\end{equation}
where the tensor $M^{\lambda\mu}$ equals
\begin{equation}
M^{\lambda\mu}=-\frac{b}{3}R\nabla^{\lambda}\phi\nabla^{\mu}\phi.
\end{equation}
Presumably, as in the case of JT theory, this expression can also be recovered from dynamical Wald entropy prescription incorporating boundary conditions~\eqref{S_W bd}. However, the actual calculation of the dynamical entropy for action~\eqref{anomaly 4D action} is too involved and we leave it for a future work.

The standard thermodynamic derivation then recovers the equations of motion
\begin{equation}
\label{anomaly 4D eom}
P_{(\mu}^{\ \ \alpha\beta\gamma}R_{\nu)\alpha\beta\gamma}-2\nabla^\alpha\nabla^\beta P_{\mu\alpha\beta\nu}+b\nabla_{(\mu}\left[\nabla_{\nu)}\phi\left(\frac{1}{6\pi}R+2\Box\phi\right)\right]-\frac{b}{3}R\nabla_{\mu}\phi\nabla_{\nu}\phi+\Psi g_{\mu\nu}=8\pi G\langle T_{\mu\nu}\rangle,
\end{equation}
where the term $b\nabla_{(\mu}\left[\nabla_{\nu)}\phi\left(R/\left(6\pi\right)+2\Box\phi\right)\right]$ comes from the Noether charge term not expressible in terms of $P^{\mu\nu\rho\sigma}$, the term $-bR\nabla_{\mu}\phi\nabla_{\nu}\phi/3$ appears through the tensor $M^{\lambda\mu}$, and the scalar function $\Psi$ is arbitrary. In the following, we list the contributions to the equations of motion from each of the terms in $P^{\mu\nu\rho\sigma}$: 

\begin{enumerate}
\item Term
\begin{align}
\nonumber \left(\frac{1}{16\pi}+\frac{a+3b}{12\pi}\phi R-\frac{b}{3}\nabla_{\lambda}\phi\nabla^{\lambda}\phi-\frac{b\Box\phi}{12\pi}+\frac{3c-2b}{288\pi^2}R\right)g^{\rho[\mu}g^{\nu]\sigma},
\end{align}
contributes
\begin{align}
\nonumber \frac{R_{\mu\nu}}{16\pi}+\left(R_{\mu\nu}+g_{\mu\nu}\Box-\nabla_{\mu}\nabla_{\nu}\right)\left(\frac{a+3b}{12\pi}\phi R-\frac{b}{3}\nabla_{\lambda}\phi\nabla^{\lambda}\phi-\frac{b\Box\phi}{12\pi}+\frac{3c-2b}{288\pi^2}R\right).
\end{align}
\item Term
\begin{align}
\nonumber \frac{\left(a+b\right)\phi}{4\pi}R^{\mu\nu\rho\sigma},
\end{align}
contributes
\begin{align}
\nonumber & \frac{\left(a+b\right)\phi}{2\pi}\left(\Box R_{\mu\nu}-\frac{1}{2}\nabla_{\mu}\nabla_{\nu}R\right)+\frac{\left(a+b\right)}{2\pi}R_{\mu\lambda\nu\rho}\nabla^{\lambda}\nabla^{\rho}\phi+\frac{\left(a+b\right)\phi}{\pi}\left(\nabla_{\lambda}R_{\mu\nu}-\nabla_{(\mu}R_{\nu)\lambda}\right)\nabla^{\lambda}\phi \\
\nonumber & +\frac{\left(a+b\right)\phi}{4\pi}\left(R_{\mu}^{\ \lambda\rho\sigma}R_{\nu\lambda\rho\sigma}+2R_{\mu\lambda\nu\rho}R^{\lambda\rho}-2R_{\mu}^{\ \lambda}R_{\nu\lambda}\right).
\end{align}
\item Term
\begin{align}
\nonumber \frac{\left(a+2b\right)\phi}{4\pi}\left(R^{\rho[\nu}g^{\mu]\sigma}+R^{\sigma[\mu}g^{\nu]\rho}\right),
\end{align}
contributes
\begin{align}
\nonumber &\frac{\left(a+2b\right)\phi}{4\pi}\left(\nabla_{\mu}\nabla_{\nu}-g_{\mu\nu}\Box\right)R-\frac{\left(a+2b\right)\phi}{4\pi}\Box R_{\mu\nu}+\frac{\left(a+2b\right)}{4\pi}\left(2R_{(\mu}^{\ \lambda}\delta^{\rho}_{\nu)}-g_{\mu\nu}R^{\lambda\rho}-R_{\mu\nu}g^{\lambda\rho}\right)\nabla_{\lambda}\nabla_{\rho}\phi \\
\nonumber & -\frac{\left(a+2b\right)\phi}{2\pi}R_{\mu\lambda\nu\rho}R^{\lambda\rho}+\frac{\left(a+2b\right)}{4\pi}\left(2\nabla_{(\nu}R_{\mu)}^{\ \lambda}-2\nabla^{\lambda}R_{\mu\nu}+\nabla_{(\mu}R\delta^{\lambda}_{\nu)}-g_{\mu\nu}\nabla^{\lambda}R\right)\nabla_{\lambda}\phi.
\end{align}
\item Term
\begin{align}
\nonumber \frac{b}{2}\left(\nabla^{\mu}\phi g^{\nu[\sigma}\nabla^{\rho]}\phi-\nabla^{\nu}\phi g^{\mu[\sigma}\nabla^{\rho]}\phi\right),
\end{align}
contributes
\begin{align}
\nonumber &bR_{\lambda(\mu}\nabla_{\nu)}\phi\nabla^{\lambda}\phi-b\nabla_{\lambda}\nabla_{\nu}\nabla_{\mu}\phi\nabla^{\lambda}\phi-b\Box\phi\nabla_{\mu}\nabla_{\nu}\phi+b\Box\nabla_{(\mu}\phi\nabla_{\nu)}\phi+\frac{b}{2}g_{\mu\nu}\nabla_{\lambda}\nabla_{\rho}\left(\nabla^{\lambda}\phi\nabla^{\rho}\phi\right).
\end{align}
\end{enumerate}
To obtain these results, we repeatedly used the second Bianchi identities. Taking the divergence of the resulting equations and imposing the local stress-energy conservation condition $\nabla^{\nu}\langle T_{\mu\nu}\rangle=0$ then both fixes the scalar function $\Psi$ to be
\begin{align}
\nonumber \Psi=&-\frac{1}{2}\mathcal{L}_{\text{GR+A}}+\frac{b}{24\pi}R\Box\phi-\frac{a+2b}{8\pi}\phi\Box R-\frac{b}{4\pi}R^{\lambda\rho}\nabla_{\lambda}\phi\nabla_{\rho}\phi \\
&-\frac{2b}{3}\Box\nabla^{\lambda}\phi\nabla^{\lambda}\phi-b\Box\phi\Box\phi+\tilde{\Lambda},
\end{align}
with $\tilde{\Lambda}$ being an arbitrary integration constant, and implies the equation of motion for the scalar field
\begin{align}
\nonumber & \Box\Box\phi-\frac{2}{3}R\Box\phi+2R^{\mu\nu}\nabla_{\mu}\nabla_{\nu}\phi+\frac{1}{3}\nabla_{\mu}R\nabla^{\mu}\phi-\frac{a+b}{8\pi b}C_{\mu\nu\rho\sigma}C^{\mu\nu\rho\sigma} \\
& -\frac{1}{12\pi}\left(R^2-3R_{\mu\nu}R^{\mu\nu}+\Box R\right)=0.
\end{align}
We thus recover the equation for $\phi$ in the same way as in the $2$D case. By taking the trace of the equations of motion for the metric and applying the equation for $\phi$, it can be easily checked that the equations of motion indeed recover the correct expression~\eqref{anomaly 4D} for the conformal anomaly of $\langle T_{\mu\nu}\rangle$. One can explicitly verify that equations~\eqref{anomaly 4D eom} agree with the equations of motion for action~\eqref{anomaly 4D action} derived in reference~\cite{Lowe:2025}.

Notably, function $\Psi$ does not correspond to the expected $-\mathcal{L}_{\text{GR+A}}/2+\tilde{\Lambda}$, but contains additional terms. These extra contributions can be linked to the presence of terms coupling curvature with second derivatives of the scalar field. Since such a situation has not been previously explicitly analyzed, the appearance of this new effect is not unreasonable. We leave its deeper analysis (possibly following similar approach as for terms containing derivatives of the curvature tensors~\cite{Mohd_2016}) for a future work.

Naturally, our derivation glosses over the irreversible entropy production terms and other effects. We present a somewhat simplified version aimed only at recovering the equations of motion. Our aim here is simply a proof of principle, i.e., showing that the effective dynamics recovering $4D$ conformal anomaly can be studied by thermodynamics methods. As argued above, a complete thermodynamic study of the $4D$ semiclassical dynamics anyway requires the proper identification of the generalized entropy. We expect to fully address the $4D$ semiclassical setting in a future work.

\section{Outlook}
\label{outlook}

The interconnection between thermodynamics and gravity has shed light on the classical structure of gravity, and as we show here, it also can open new paths in the comprehension of gravitational theories beyond the classical realm. Specifically, we dug in the search of understanding quantum effects in gravity by the introduction of the well-known quantum character of matter and consider the framework of semi-classical gravity (where the backreaction of the quantum matter onto the spacetime geometry) derived from thermodynamics of spacetime. 

For this purpose, we have first developed a framework for deriving the equations of motion of scalar-tensor theories of gravity from thermodynamics. Our approach uses the previously introduced augmented Wald entropy, but offers its physical interpretation in terms of the Wald entropy of a quasi-stationary horizon with Neumann boundary conditions imposed on the scalar field, explicitly shown in $2$D. We further proposed that the scalar field potential enters the equations of motion as a work performed by an external force imposed on the stretched light cone by the exterior spacetime. Using these tools, we then focused on rigorously deriving the semi-classical gravitational field equation in $2$-dimensional JT gravity from generalized entropy of locally constructed stretched light cones, incorporating conformal quantum matter fields and their corresponding backreaction. As a side product, we have also shown that the dynamical Wald entropy agrees with the generalized entropy, improving previous results obtained in the context of black hole thermodynamics in JT gravity.

We have finally extended our analysis to $4$-dimensional general relativity, for which the semi-classical backreaction does not appear to be fully encoded in the conformal anomaly as in the previous $2$-dimensional case. In this context we have been able to analyze the derivation of the gravitational dynamics recovering the conformal anomaly from the augmented Wald entropy. The future perspective involves, first, to compute the dynamical entropy in this setup and compare it with the augmented Wald entropy; and second, to extend the analysis to the (presently unknown) full generalized entropy and complete the semi-classical dynamical picture of $4$-dimensional gravity.

In addition, and beyond the scope of this paper, the introduction of the augmented Wald entropy poses an interesting question to be tackled in future work. The fact that the augmented Wald entropy agrees with the dynamical prescription offers a valuable gravitational interpretation of the extra ambiguity-based terms. However, the statistical interpretation remains an open question. The subtlety of this problem comes from the fact that strong coupling prohibits a clear separation between (quantum) matter and gravity, which is generally a problematic aspect of strongly coupled semi-classical gravity. However, in the case of semi-classical JT gravity, this problem might be solvable using the framework of von Neumann algebras. They have already been used to characterise generalised entropy in gravitational settings in the weak coupling regime, e.g. in the de Sitter patch \cite{Chandrasekaran_2023}, and even generalised to arbitrary subregions \cite{Jensen_2023}. The type and properties of von Neumann algebra of JT gravity has also been investigated \cite{Harlow_2022}, and there is hope that previous considerations of generalised entropy could be applied in this regime to explain the appearance of the extra terms. Note that resolving this issue in the case of two-dimensional semi-classical gravity could shed some light on the microscopic nature of entropy in more general scalar-tensor theories.

\section*{Acknowledgments}

The authors thank Andrew Svesko for contributing to the initial idea of the project and for numerous useful discussions and comments on the manuscript. The authors further acknowledge helpful and constructive comments from an anonymous referee. AA-S is funded by the Deutsche Forschungsgemeinschaft (DFG, German Research Foundation) — Project ID 51673086. ML is supported by the DIAS Post-Doctoral Scholarship in Theoretical Physics 2024. AA-S also acknowledges partial support through Grant  No.  PID2023-149018NB-C44 (funded by MCIN/AEI/10.13039/501100011033). This work was supported by the Engineering and Physical Sciences Research Council [grant number EP/S021582/1].

\appendix

\section{Wald entropy}
\label{Wald entropy review}

We introduce the covariant phase space formalism for a $D$-dimensional spacetime $\left(\mathcal{M},g_{\mu\nu}\right)$ with a boundary $\partial\mathcal{M}$, consisting of an initial spacelike surface $\Sigma_{\text{i}}$, a final spacelike surface $\Sigma_{\text{f}}$, and a lateral timelike boundary $\Gamma$ connecting these surfaces. We denote the intersections of $\Gamma$ with $\Sigma_{\text{i}}$ and $\Sigma_{\text{f}}$ by $\partial\Sigma_{\text{i}}$ and $\partial\Sigma_{\text{f}}$, respectively.

On this manifold, consider a gravitational theory with a Lagrangian $D$-form $\boldsymbol{L}_{\text{bulk}}$ and a boundary Lagrangian $\left(D-1\right)$-forms $\boldsymbol{L}_{\text{i}}$, $\boldsymbol{L}_{\text{f}}$, $\boldsymbol{L}_{\Gamma}$ and $\boldsymbol{L}_{\mathcal{H}}$. A small, arbitrary variation of $\boldsymbol{L}_{\text{bulk}}$  yields
\be
\boldsymbol{L}_{\text{bulk}}=\boldsymbol{E}_{\psi}\delta\psi+\text{d}\boldsymbol{\theta}_{\text{bulk}}\left[\delta\right],
\ee
where we use $\psi$ to collectively represent the dynamical fields. The equations of motion read $\boldsymbol{E}_{\psi}=0$. The $\left(D-1\right)$-form $\boldsymbol{\theta}_{\text{bulk}}\left[\delta\right]$ is then the bulk symplectic potential. Similarly, varying the total boundary Lagrangian $\boldsymbol{L}_{\text{bd}}=\boldsymbol{L}_{\text{i}}+\boldsymbol{L}_{\text{f}}+\boldsymbol{L}_{\Gamma}+\boldsymbol{L}_{\mathcal{H}}$ leads to
\be
\delta\boldsymbol{L}_{\text{bd}}=-j^{*}_{\text{bd}}\boldsymbol{\theta}_{\text{bulk}}\left[\delta\right]+\boldsymbol{B}\left[\delta\right]+\text{d}\boldsymbol{\theta}_{\text{bd}}\left[\delta\right].
\ee
Here, $j^{*}_{\text{bd}}$ denotes the pullback to the boundary, $\left(D-1\right)$-form $\boldsymbol{B}\left[\delta\right]$ vanishes when the boundary conditions associated with $\boldsymbol{L}_{\text{bd}}$ are satisfied, and the $\left(D-2\right)$-form $\boldsymbol{\theta}_{\text{bd}}\left[\delta\right]$ is the boundary symplectic potential. For the symplectic form $\Omega\left[\delta_1,\delta_2\right]$ on a spacelike Cauchy surface $\Sigma$ evaluated for two independent variations $\delta_1$, $\delta_2$ we have
\be
\Omega\left[\delta_1,\delta_2\right]=\int_{\Sigma}\left[\delta_1\boldsymbol{\theta}_{\text{bulk}}\left[\delta_2\right]-\delta_1\boldsymbol{\theta}_{\text{bulk}}\left[\delta_2\right]-\text{d}\left(\delta_1\boldsymbol{\theta}_{\text{bd}}\left[\delta_2\right]-\delta_1\boldsymbol{\theta}_{\text{bd}}\left[\delta_2\right]\right)\right].
\ee
It is easy to check that $\Omega\left[\delta_1,\delta_2\right]$ depends on $L_{\text{bd}}$ only through $\boldsymbol{B}\left[\delta\right]$. If the boundary conditions are satisfied, i.e., $\boldsymbol{B}\left[\delta\right]=0$, $\Omega\left[\delta_1,\delta_2\right]$ only depends on the bulk Lagrangian~\cite{Neri:2024}. Similarly, $\Omega\left[\delta_1,\delta_2\right]$ remains invariant under shifting $\boldsymbol{\theta}_{\text{bulk}}\left[\delta\right]$ by an arbitrary exact  $\left(D-1\right)$-form $\text{d}\boldsymbol{Y}$.

For a variation corresponding to an infinitesimal diffeomorphism generated by vector field $\xi$. The corresponding bulk Noether current conserved on shell than reads
\be
\boldsymbol{j}_{\text{bulk}}\left[\pounds_{\xi}\right]=\boldsymbol{\theta}_{\text{bulk}}\left[\pounds_{\xi}\right]-\iota_{\xi}\boldsymbol{L}_{\text{bulk}}, \label{current bulk}
\ee
where $\iota_{\xi}$ is the interior product with $\xi$. Similarly, the boundary Noether current reads
\be
\boldsymbol{j}_{\text{bd}}\left[\pounds_{\xi}\right]=\boldsymbol{\theta}_{\text{bd}}\left[\pounds_{\xi}\right]-\iota_{\xi}\boldsymbol{L}_{\text{bd}}.
\ee
It is conserved if the boundary conditions are satisfied, i.e., $\boldsymbol{B}\left[\delta\right]=0$. The diffeomorphism Noether charge associated with a spacelike Cauchy surface $\Sigma$ and conserved on shell then obeys
\be
\label{charge}
Q\left[\pounds{L}_{\xi}\right]=\int_{\Sigma}\boldsymbol{j}_{\text{bulk}}\left[\pounds{L}_{\xi}\right]-\int_{\partial\Sigma}\boldsymbol{j}_{\text{bd}}\left[\pounds{L}_{\xi}\right].
\ee
Once we choose the boundary Lagrangian, the corresponding Noether charge is unique~\cite{Harlow_2020,Margalef:2021,Rignon:2023,Neri:2024}. Notably, the definition remains unique in fully dynamical situations, although it then depends on the choice of the spacelike Cauchy surface $\mathcal{C}$.

A variation of the Noether charge expression evaluated on shell yields
\be
\label{first law d}
\delta Q\left[\pounds{L}_{\xi}\right]=\Omega\left[\delta,\pounds{L}_{\xi}\right].
\ee
Evaluating this relation for a black hole spacetime provides the variational first law of black hole thermodynamics. Similarly, the conservation of the Noether charge between the initial Cauchy surface $\Sigma_{\text{i}}$ and the final Cauchy surface $\Sigma_{\text{f}}$,
\be
\label{first law p}
Q\left[\pounds{L}_{\xi}\right]\vert_{\Sigma_{\text{f}}}-Q\left[\pounds{L}_{\xi}\right]\vert_{\Sigma_{\text{i}}}=0,
\ee
encodes the physical process first law of black hole thermodynamics.

This allows us to define black hole entropy as the total contribution to the Noether charge on the black hole horizon divided by the Hawking temperature. Of course, in highly dynamical situations, the definition of the horizon itself becomes ambiguous. Nevertheless, a choice of suitable boundary conditions allows a meaningful definition of entropy at least in quasi-stationary situations, such as on the surface of a stretched light cone. In particular, imposing conformal boundary conditions for the metric on the perturbed Killing horizon~\cite{Rignon:2023} reproduce the recently proposed dynamical entropy prescription~\cite{Hollands_2024,Visser_2024}. In general relativity, both prescriptions recover entropy proportional to the area of the apparent horizon.

\bibliography{stthermorefs}

\end{document}